\documentclass[journal]{IEEEtran}
\usepackage{cite}

\ifCLASSINFOpdf
   \usepackage[pdftex]{graphicx}
  \graphicspath{{figs/}}
  \DeclareGraphicsExtensions{.pdf,.jpeg,.png}
\else
  \usepackage[dvips]{graphicx}
  \graphicspath{{figs/}}
  \DeclareGraphicsExtensions{.eps}
\fi
\usepackage{amsmath,amssymb,amsthm,amsfonts,amscd,amsbsy,amsxtra,amsfonts}
\usepackage{mathtools}
\DeclareMathAlphabet{\mathpzc}{OT1}{pzc}{m}{it} 
\usepackage[hyphens]{url}

\usepackage[dvipsnames]{xcolor}
\usepackage[normalem]{ulem}
\usepackage{array, makecell, cellspace}
\usepackage{enumitem}

\usepackage{tablefootnote}

\DeclareMathOperator{\sinc}{Sinc}
\DeclareMathOperator{\rect}{\Pi}
\DeclarePairedDelimiter\ceil{\lceil}{\rceil}

\newtheorem{prop}{Proposition}

\usepackage{acro}
\usepackage{multicol}

\newlist{acronyms}{description}{1}
\setlist[acronyms]{
  labelwidth = 6em,
  leftmargin = 7em,
  noitemsep,
  itemindent = 0pt
}

\NewAcroTemplate[list]{twocolumn}{%
  \begin{multicols}{2}[\acroheading]
    \acropreamble
    \begin{acronyms}
      \acronymsmapF
        {\item[\bfseries\acrowrite{short}] \acrowrite{list}}
        {\item\AcroRerun}
    \end{acronyms}
  \end{multicols}
}

\DeclareAcronym{ltv}{short=LTV,long=linear time-varying}
\DeclareAcronym{lti}{short=LTI,long=linear time-invariant}
\DeclareAcronym{tvcir}{short=TV-CIR,long=time-varying channel impulse response}
\DeclareAcronym{td}{short=TD,long=time-domain}
\DeclareAcronym{fd}{short=FD,long=frequency-domain}
\DeclareAcronym{dd}{short=DD,long=delay-Doppler}
\DeclareAcronym{tf}{short=TF,long=time-frequency}
\DeclareAcronym{mc}{short=MC,long=multi-carrier}
\DeclareAcronym{sc}{short=SC,long=single-carrier}
\DeclareAcronym{ofdm}{short=OFDM,long=orthogonal frequency division multiplexing}
\DeclareAcronym{oddm}{short=ODDM,long=orthogonal delay-Doppler  division multiplexing}
\DeclareAcronym{otfs}{short=OTFS,long=orthogonal time frequency space}
\DeclareAcronym{ddop}{short=DDOP,long=delay-Doppler domain orthogonal pulse}
\DeclareAcronym{tfop}{short=TFOP,long=time-frequency domain orthogonal pulse}
\DeclareAcronym{ddlp}{short=DDLP,long=delay-Doppler domain localized pulse}
\DeclareAcronym{ddmc}{short=DDMC,long=delay-Doppler domain multi-carrier}
\DeclareAcronym{jtfr}{short=JTFR,long=joint time-frequency resolution}
\DeclareAcronym{stft}{short=STFT,long=short-time Fourier transform}
\DeclareAcronym{wh}{short=WH, long= Weyl-Heisenberg}
\DeclareAcronym{isfft}{short=ISFFT, long=inverse symplectic finite Fourier transform}
\DeclareAcronym{fft}{short=FFT, long=fast Fourier transform}
\DeclareAcronym{dft}{short=DFT, long=discrete Fourier transform}
\DeclareAcronym{idft}{short=IDFT, long=inverse discrete Fourier transform}
\DeclareAcronym{ifft}{short=IFFT, long=inverse fast Fourier transform}
\DeclareAcronym{tfmc}{short=TFMC,long=time-frequency domain multi-carrier}
\DeclareAcronym{esdd}{short=ESDD,long=equivalent sampled delay-Doppler}
\DeclareAcronym{io}{short=IO,long=input-output}
\DeclareAcronym{isac}{short=ISAC,long=integrated sensing and communications}
\DeclareAcronym{dmt}{short=DMT,long=discrete multi-tone}
\DeclareAcronym{pa}{short=PA,long=power amplifier}
\DeclareAcronym{bpf}{short=BPF,long=bandpass filter}
\DeclareAcronym{lna}{short=LNA,long=low-noise amplifier}
\DeclareAcronym{lpf}{short=LPF,long=low-pass filter}
\DeclareAcronym{oobe}{short=OOBE,long=out-of-band emission}
\DeclareAcronym{rf}{short=RF,long=radio frequency}
\DeclareAcronym{dsp}{short=DSP,long=digital signal processor}
\DeclareAcronym{isi}{short=ISI,long=inter-symbol-interference}
\DeclareAcronym{ici}{short=ICI,long=inter-carrier-interference}
\DeclareAcronym{qam}{short=QAM,long=quadrature amplitude modulation}
\DeclareAcronym{tfa}{short=TFA,long=time-frequency area}
\DeclareAcronym{gsm}{short=GSM,long=global system of mobile}
\DeclareAcronym{dof}{short=DoF,long=degree of freedom}
\DeclareAcronym{1d}{short=1D,long=one-dimensional}
\DeclareAcronym{2d}{short=2D,long=two-dimensional}
\DeclareAcronym{iq}{short=IQ,long=in-phase and quadrature}
\DeclareAcronym{dsl}{short=DSL,long=digital subscriber line} 
\DeclareAcronym{cp}{short=CP,long=cyclic prefix}
\DeclareAcronym{cs}{short=CS,long=cyclic suffix}
\DeclareAcronym{tlop}{short=TLOP,long=time-limited orthogonal pulse}
\DeclareAcronym{blop}{short=BLOP,long=band-limited orthogonal pulse}
\DeclareAcronym{fbmc}{short=FBMC,long=filter bank multi-carrier}
\DeclareAcronym{psofdm}{short=PS-OFDM,long=pulse-shaped OFDM}
\DeclareAcronym{ptsofdm}{short=PTS-OFDM,long=pulse-train-shaped OFDM}
\DeclareAcronym{oqam}{short=OQAM,long=offset quadrature amplitude modulation}
\DeclareAcronym{smt}{short=SMT,long=staggered multi-tone}
\DeclareAcronym{tmx}{short=TMX,long=time division multiplexer}
\DeclareAcronym{dac}{short=DAC,long=digital-to-analog converter}
\DeclareAcronym{cfo}{short=CFO,long=carrier frequency offset}
\DeclareAcronym{ber}{short=BER,long=bit error rate}
\DeclareAcronym{psw}{short=PSW,long=prolate spheroidal wave}
\DeclareAcronym{tdm}{short=TDM,long=time-division multiplexing}
\DeclareAcronym{rrc}{short=RRC,long=root raised cosine}
\DeclareAcronym{eva}{short=EVA,long= Extended Vehicular A}
\DeclareAcronym{dfrc}{short=DFRC,long=dual-functional radar-communication}
\DeclareAcronym{isll}{short=ISLL,long=integrated side-lobe level}
\DeclareAcronym{sisll}{short=SISLL,long=sampled ISLL}
\DeclareAcronym{2g}{short=2G,long=the second generation mobile communication system}
\DeclareAcronym{3g}{short=3G,long=the third generation mobile communication system}
\DeclareAcronym{4g}{short=4G,long=the fourth generation mobile communication system}
\DeclareAcronym{5g}{short=5G,long=the fifth generation mobile communication system}
\DeclareAcronym{6g}{short=6G,long=the sixth generation mobile communication system}
\DeclareAcronym{psd}{short=PSD,long=power spectral density}
\DeclareAcronym{nmse}{short=NMSE,long=normalized mean square error}
\DeclareAcronym{ue}{short=UE,long=user equipment}
\DeclareAcronym{cdma}{short=CDMA,long= code-division multiple access}
\DeclareAcronym{ofdm-im}{short=OFDM-IM,long= orthogonal frequency division multiplexing combined with index modulation}
\DeclareAcronym{im}{short=IM,long=index modulation}
\DeclareAcronym{dft-s-ofdm}{short=DFT-S-OFDM, long= discrete Fourier transform-spread-orthogonal frequency division multiplexing}
\DeclareAcronym{mimo}{short=MIMO,long=multiple-input multiple-output}
\DeclareAcronym{miso}{short=MISO,long=multiple-input single-output}
\DeclareAcronym{papr}{short=PAPR, long=peak-to-average power ratio} 
\DeclareAcronym{rsma}{short=RSMA, long= rate-splitting multiple access}
\DeclareAcronym{noma}{short=NOMA, long=non-orthogonal multiple access}
\DeclareAcronym{dc}{short=DC, long=direct-current}
\DeclareAcronym{ocdm}{short=OCDM, long=orthogonal chirp division multiplexing}
\DeclareAcronym{afdm}{short=AFDM, long=affine frequency division multiplexing}

\begin{document}
\bstctlcite{IEEEexample:BSTcontrol}
%

\title{Multi-Carrier Modulation: An Evolution from Time-Frequency Domain to Delay-Doppler Domain}
\IEEEspecialpapernotice{(Invited Paper)}

%
%
%

\author{Hai~Lin,~\IEEEmembership{Senior Member,~IEEE,}
  Jinhong~Yuan,~\IEEEmembership{Fellow,~IEEE,}
  Wei~Yu,~\IEEEmembership{Fellow,~IEEE,} \\
  Jingxian~Wu,~\IEEEmembership{Fellow,~IEEE,}   Lajos~Hanzo,~\IEEEmembership{Life Fellow,~IEEE}
  \thanks{This paper has been accepted for publication in IEEE Trans. Commun. It was presented in part at the IEEE GLOBECOM 2022, Rio de Janeiro, Brazil\cite{ddop}. Further  information on ODDM modulation is available at \protect\url{https://oddm.io}} 
  \thanks{H. Lin is with the Department of Electrical and Electronic Systems Engineering, Graduate School of Engineering, Osaka Metropolitan University, Sakai, Osaka 599-8531, Japan (e-mail: hai.lin@ieee.org).}
  \thanks{J. Yuan is with the School of Electrical Engineering and Telecommunications, the University of New South Wales, Sydney, Australia (e-mail: j.yuan@unsw.edu.au).}%
  \thanks{W. Yu is with the Edward S. Rogers Sr. Department of Electrical and Computer Engineering, the University of Toronto, Toronto, Canada (e-mail: weiyu@comm.utoronto.ca).}%
  \thanks{J. Wu is with the Department of Electrical Engineering, University of Arkansas, Fayetteville, Arkansas, USA (e-mail: wuj@uark.edu).}
  \thanks{L. Hanzo is with the School of Electronics and Computer Science, University of Southampton, Southampton SO17 1BJ, U.K (e-mail:
    lh@ecs.soton.ac.uk).}}

\maketitle


\begin{abstract}
  The recently proposed orthogonal delay-Doppler division multiplexing (ODDM) modulation, which is a delay-Doppler (DD) domain multi-carrier (DDMC) modulation scheme based on the DD domain orthogonal pulse (DDOP), is studied.
  We first revisit the linear time-varying (LTV) channel model for the wireless channel, and review the conventional multi-carrier (MC) modulation schemes and their design guidelines for both linear time-invariant (LTI) and LTV channels. We then focus on the representation of the LTV channel in an equivalent sampled DD (ESDD) domain, and propose an impulse-function-based transmission strategy for the ESDD channel. Next, we take an in-depth look into the DDOP and show that it achieves orthogonality with respect to the fine time and frequency resolutions in the ESDD domain thus \emph{behaves like} an impulse function. This allows us to unveil the unique input-output relation of the resultant ODDM modulation over the ESDD channel. We point out that the conventional MC modulation design guidelines based on the Weyl-Heisenberg (WH) frame theory can be relaxed without compromising its orthogonality or violating the WH frame theory. More specifically, for a practical communication system with bandwidth and duration constraints, MC modulation signals can be designed considering so-called \emph{local or sufficient (bi)orthogonality}, which refers to the (bi)orthogonality among a WH \emph{subset} for the MC signal within a specific bandwidth and duration.
  This is different from the conventional MC modulation waveform design guidelines (such as for orthogonal frequency division multiplexing and orthogonal time frequency space) based on the global (bi)orthogonality, which is the (bi)orthogonality among a WH \emph{full set} corresponding to the MC signal occupying the entire TF domain.
  This novel design guideline could potentially open up opportunities for developing future waveforms required by new applications such as communication systems associated with high delay and/or Doppler shifts, as well as integrated sensing and communications.
\end{abstract}


%

\IEEEpeerreviewmaketitle

\begin{figure*}
  \printacronyms[template=twocolumn]
\end{figure*}


\section{Introduction}
%
%
%
%

A wireless channel typically introduces both time and frequency dispersions, which correspond to the channel's frequency and time selectivity, respectively. Usually, such a doubly-selective channel can be modeled as a \ac{ltv} system, and it is represented by its \ac{tvcir} or \ac{dd} spread function\cite{bello}, a.k.a. the spreading function\cite{Hlawatsch2011}.
Within the channel's \emph{coherence time}, the channel model can be simplified to a \ac{lti} system, which only has time dispersion.
The channel-induced dispersions have a crucial impact on signal transmission and therefore become the primary concern in the design of modulation schemes.

In digital communications, a modulation technique is essentially a scheme of using \emph{analog pulses} or mathematically equivalent \emph{continuous-time functions} to synthesize transmit signal waveforms, where each pulse carries an information-bearing \emph{digital symbol} drawn from a signal constellation diagram\cite{Jacobs_pce}. In other words, a ``symbol” is represented by a continuous-time function, which usually is the product of a digital symbol and an analog pulse. At the receiver, demodulation is often performed first by receive filtering based on matched filters or correlators corresponding to these analog pulses. Then, the extracted signal components are fed into a channel equalizer to recover the transmitted digital symbols. To avoid interference among the symbols and consequently ease the channel equalization, it is expected that these analog pulses do not interfere with each other, if possible, even in the presence of channel dispersions.
As a result, (bi)orthogonal pulses or functions, which prevent mutual interference upon obeying the (bi)orthogonality among themselves, lie at the \emph{very heart} of modulation techniques.

Since the eigenfunctions of a linear system are orthogonal functions and excite simple scaled system outputs, using eigenfunctions as the pulses for signal transmission seems to be an ideal strategy.
For LTI systems, the eigenfunctions are complex sinusoids with \emph{infinite} duration \cite{sigsyt}, while any practical pulse must have a \emph{finite} duration. Fortunately, complex sinusoids are \emph{periodic} functions, which implies that if we appropriately choose their frequencies, truncated complex sinusoids can still be orthogonal to each other and exhibit a \emph{scalar} system \ac{io} relation.
The corresponding modulation scheme conceived for LTI channels is the popular \ac{tf} domain \ac{mc} modulation, typically, the \ac{ofdm}\cite{fwc,ofdm,tff,mct}\footnote{MC modulation is a general term, while the OFDM is a special form of MC modulation\cite{ofdm}.}, which has been widely adopted in wireless standards, such as the Wi-Fi\cite{80211a,80211ac}, \ac{4g}\cite{4Gbook}, \ac{5g}\cite{5Gbook}, and so on.

In OFDM, to achieve eigenfunction-based transmission, the transmit pulses are chosen as complex sinusoids, also known as \emph{subcarriers} or \emph{tones}, and then truncated by a common \ac{td} window function called \emph{prototype pulse}\cite{tff}.
Therefore, a popular way of thinking about the transmit pulses in the MC modulation schemes is to treat them as TF-shifted prototype pulses along a TF grid.
In fact, MC modulation can be defined by its prototype pulse and TF grid parameterized by a specific frequency resolution (subcarrier spacing) $\Delta F$ and a specific time resolution (symbol interval) $\Delta T$, where the \ac{jtfr} is given as $\Delta R= \Delta T \Delta F$.

\begin{figure}
  \centering
  \includegraphics[width=7.5cm]{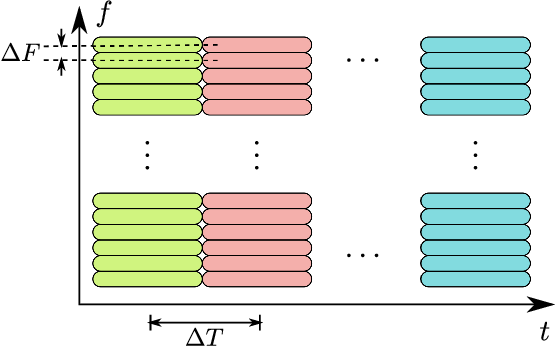}
  \caption{TF grid in MC modulation}
  \label{tf_grid}
\end{figure}

An example of the TF grid in MC modulation is shown in Fig. \ref{tf_grid}, where each color block presents a transmit pulse obtained by TF-shifting the prototype pulse.
Given $\Delta T$ and $\Delta F$, the \ac{2d} TF domain in Fig. \ref{tf_grid} is a \emph{gridded} \ac{2d} domain \emph{dedicated for} representing the discrete TF structure and the signal (energy) localization of the MC modulation.
Then, the fundamental issue of designing an MC modulation scheme becomes that of finding the (bi)orthogonal prototype pulse with respect to $\Delta T$ and $\Delta F$. Traditionally, the transmit pulses in Fig. \ref{tf_grid} are considered as a \emph{\ac{wh}} or \emph{Gabor} function set\cite{tff}, which is the principal tool of the \ac{stft} based \ac{tf} analysis\cite{wavelet,gaborana,ftfa}.
According to the WH frame theory, the (bi)orthogonal WH function sets only exist for $\Delta R \ge 1$\cite{wexler1990,janssen1995}, and therefore most MC modulation schemes are designed for $\Delta R \ge 1$ \cite{tff,mct}.

The situation becomes more complicated when the channel's frequency dispersion is significant, for example, for an LTV channel encountered in a high-mobility environment.
In contrast to LTI channels, which have elegant and \emph{common} eigenfunctions to ease the pulse design, the underspread LTV channels at best have a structured set of \emph{approximate} eigenfunctions, which depend on the spreading or scattering function of the channel\cite{eigen}.
In fact, these approximate eigenfunctions require not only a channel-dependent prototype pulse but also a much coarser JTFR than that of the classic OFDM \cite{eigen,matz07}.
The channel-dependency and the coarse JTFR make it impracticable to realize eigenfunction-based transmission over LTV channels.

Recently, \ac{dd} domain modulation schemes, including the popular \ac{otfs} \cite{otfs_wcnc_2017,hadani_otfs_2018} and the \ac{oddm}\cite{oddmicc22,oddm}, have been proposed to address the challenges of waveform design for LTV or doubly-selective channels.
The rationale behind the DD domain modulation schemes is to explore the fact that LTV channels often have a much longer \emph{stationary time}\footnote{Although the term of stationary time was originally used in the stochastic modelling of LTV channels, here we refer to it as the time span over which the channel's delay and Doppler shifts remain almost the same.} over which the DD state of the channel remains approximately constant, as compared to their \emph{coherence time} over which the TF state is approximately constant\cite{Hlawatsch2011}.  Although the channel's TV-CIR still needs a large number of parameters within the stationary time, its DD spread function becomes \emph{deterministic} and has a compact representation relying on much fewer parameters.
Since each propagation path imposes not only attenuation and delay but also the Doppler spread, the TV-CIR of an LTV channel is given by the superposition of path-wise complex sinusoids associated with Doppler frequencies, and therefore presents \emph{fading}.
On the other hand, its DD spread function is simply characterized by each path's attenuation, delay and Doppler. Then, the destructive multi-path fading exhibited by the TV-CIR turns into separable paths in the DD domain, where path diversity can be harvested because the paths are independent\cite{yuan_wcm_2021}.
Based on these observations, the DD domain modulation schemes aim for \emph{coupling} \cite{hadani_otfs_2018} the modulated signal with the DD domain channel or equivalently with the DD spread function of the LTV channel.
Here, coupling represents the match between the TF grid namely the TF resolutions of the modulated signal and those of the DD domain channel\cite{oddm}.
Given an ideal coupling, the stability and sparsity of DD domain channels can be exploited to obtain diversity gains with \emph{minimum} interference, and hence achieve reliable communications with low pilot overhead and low processing complexity\cite{hadani_otfs_2018}.

As for any modulation scheme, the fundamental issue of a DD domain modulation also resides in the pulse design.
Intuitively, to achieve the signal-channel coupling with matched resolutions, the DD domain modulation requires a localized pulse in the DD domain. However, such a \ac{ddlp} would violate the \emph{Heisenberg uncertainty principle} \cite{hadaniyt}, and therefore does not exist.
To circumvent this issue, OTFS first converts the digital DD domain symbols to the digital TF domain signals via an \ac{isfft} based \emph{digital precoder}.
Then, realizable orthogonal pulses in the TF domain are employed to carry these precoded signals and synthesize a conventional \ac{tfmc} modulation waveform, typically OFDM\cite{otfs_wcnc_2017}.
As a result, the transmit pulses of OTFS are still the \ac{tfop}, and
the OTFS is essentially a precoded OFDM\cite{zemenpimrc2018}.
In OTFS, the \emph{ideal pulse assumed has to} be \emph{bi-orthogonal robust} against the channel-induced delay and Doppler. Unfortunately, such an ideal pulse does not exist either\cite{hadani_otfs_2018}, as none of the TFOPs can meet this requirement.
Meanwhile, the widely adopted rectangular prototype pulse of popular OTFS studies is still a TFOP and therefore non-ideal, which will face practical challenges in implementation, such as high \ac{oobe} and severe interference\cite{shen2022error}.

On the other hand, based on the newly discovered \ac{ddop}, ODDM represents a novel \ac{ddmc} modulation scheme that can avoid the impediments of OTFS mentioned above\cite{oddm}.
Note that the time (duration) and frequency (bandwidth) constraints of any practical waveform define a TF region of interests, which results in
\begin{enumerate}
  \item an \ac{esdd} channel model with its delay and Doppler resolutions\cite{bello};
  \item a corresponding gridded DD domain.
\end{enumerate}
It is noteworthy that the gridded DD domain is essentially a gridded TF domain but with much finer resolutions, {because the delay and Doppler have \emph{the physical units} of time and frequency, respectively. Thus, considering the presence of frequency resolution, a DD domain modulation is naturally a multi-carrier modulation, which requires a DDLP or a DDOP.
Although the DDLP does not exist,
the DDOP introduced in \cite{oddmicc22,oddm,ddop} consists of a train of square-root Nyquist pulses, and \emph{behaves like} the ``nonexistent" DDLP in the TF region of interests,  and it achieves perfect coupling between the modulated signal and the ESDD channel.
Specifically, without violating the Heisenberg uncertainty principle, the DDOP has an equally-spaced signal localization in the TF region of interests\cite{ddop} to satisfy the orthogonality with respect to the specific delay and Doppler resolutions.
It should be noted that the ESDD channel actually obeys the classic equivalent sampled channel model\cite{fwc,bello}, which is the \emph{on-grid} equivalent of the \emph{effective} channel corresponding to the cascaded transmit filter, propagation channel, and receive filter. Naturally, it is the ESDD channel that really matters for transceiver design.

In comparison to the conventional TFMC modulation schemes associated with $\Delta R \ge 1$, the DDOP-based ODDM or general DDMC modulation is a new type of MC modulation having a much reduced JTFR of $\Delta R \ll 1$, as we show later. The ambitious objective of this paper is to analytically appraise this new modulation.

We commence by revisiting the LTV and LTI channels, and then review the conventional TFMC modulation schemes in terms of their transmission strategy, pulse design principles, implementation methods, and performance over LTV channels. Then we discuss MC modulation schemes designed for the ESDD channel.
We characterize the \emph{time-varying property} of the ESDD channel, and propose an impulse-function-based transmission strategy for it. Next, we take an in-depth look into the DDOP and show that it achieves orthogonality with respect to the fine time and frequency resolutions in the ESDD domain thus behaves like an impulse function. This allows us to unveil the unique input-output relation of the
resultant ODDM modulation over the ESDD channel. Then, we point out that the conventional MC modulation design guidelines based on the WH frame theory can be relaxed without compromising its orthogonality. 
In particular, instead of the global (bi)orthogonality governed by the WH frame theory, the MC pulse design can be redefined by exploiting the WH \emph{subset} based \emph{local or sufficient (bi)orthogonality}.
This new interpretation of the local/sufficient (bi)orthogonality actually relaxes the JTFR constraint of $\Delta R\ge 1$ for (bi)orthogonal pulse design and leads to more general DDMC modulation schemes. This novel design guideline may open up opportunities for developing future waveforms required by new applications such as, \ac{isac}, high-mobility communications, etc.

The rest of the paper is organized as follows:
Section II revisits the LTV and LTI channel models, especially the ESDD channel model taking into account the time and frequency constraints of practical signal waveforms. The family of classic TFMC modulation schemes is reviewed in Section III and Section IV for LTI channels and LTV channels, respectively, focusing on their transmission strategies, pulse designs and implementation methods. Then, based on the continuous-time channel IO relation, Section V investigates the properties of the ESDD channel, clarifies the corresponding time-varying DD domain impulse response, and proposes an impulse-function-based  transmission strategy. The DDOP and its (bi)orthogonality are analyzed in detail in Section VI, where the relation to the WH frame theory is explained and new pulse design guidelines are proposed. The important properties of the ODDM modulation are unveiled in Section VII, including its signal localization, bandwidth efficiency, implementation methods, and ISAC potentials. Our simulation results are provided in Section VIII and finally Section IX concludes the paper.

\textbf{Notations}: In this paper, uppercase boldface letters are used to represent matrices, and lowercase boldface letters are used for column vectors. Furthermore, $\rect_ T(t)$ denotes the rectangular function of unit energy and TD support of $[0, T]$, and $\sinc(\chi) \triangleq \frac{\sin(\pi \chi)}{\pi \chi}$.
The superscript $T$ denotes the transpose operator, while $[\cdot]_M$ stands for the mod $M$ operator.
Finally, $\mathcal A_{g,\tilde g}(\tau,\nu)$ is the cross ambiguity function between two pulses $g(t)$ and $\tilde g(t)$, given by
\begin{align*}
  \mathcal A_{g,{\tilde g}}(\tau,\nu) & =   \langle g(t), {\tilde g}(t-\tau)e^{j2\pi \nu (t-\tau)} \rangle,                          \\
                                      & =  \int_{-\infty}^{\infty} g(t) {\tilde g}^{*} (t-\tau)e^{-j2\pi \nu (t-\tau)} dt. \nonumber
\end{align*}

\section{Wireless Channel Models}

\subsection{Propagation Channel Models}
Consider a complex-valued baseband signal $x(t)$ in the band of $[-B/2, B/2]$. Due to its limited bandwidth, $x(t)$ cannot be \emph{strictly} time-limited. Similarly, a time-limited signal cannot be \emph{strictly} frequency-limited. To accommodate practical time-limited signals having finite duration, the bandwidth here is defined in an essential sense of \cite{onbandwidth}\footnote{The bandwidth may be defined by ignoring the negligibly small high-frequency tails beyond $[-B/2, B/2]$.}. Then, $x(t)$ having a duration of $T_x$ may be considered as both time- and band-limited.

For a wireless system communicating over an LTV channel, given the carrier frequency $f_c$, the passband \ac{rf} signal $\mathsf x(t)=\Re\{x(t)e^{j2\pi f_c t}\}$ is amplified and then sent through the LTV channel. We assume that the LTV channel is composed of $\tilde P$ paths corresponding to $\tilde P$ discrete specular scatterers. Under the ``narrowband" assumption of $B\ll f_c$,  the variation of path attenuations and propagation delays versus frequency can be omitted. Furthermore, noise terms are ignored in the following discussion for the sake of simplicity. Then, we have the received \emph{real-valued passband signal}\cite{fwc}
\begin{equation}
  \mathsf y(t)=\sum_{p=1}^{\tilde P} {\mathsf a}_p(t)\mathsf x(t-\tilde \tau_p(t)),
\end{equation}
where ${\mathsf a}_p(t)$ and $\tilde \tau_p(t)$ are the time-varying attenuation and delay of the $p$-th path, respectively.
The corresponding received \emph{complex-valued baseband signal} is
\begin{equation}\label{bbyt_general}
  y(t)=\sum_{p=1}^{\tilde P}\tilde h_p(t) x(t-\tilde \tau_p(t)),
\end{equation}
where $\tilde h_p(t)={\mathsf a}_p(t)e^{-j2\pi f_c \tilde \tau_p(t)}$ represents the ``gain" or the attenuation
of the $p$-th path.
Thus, the baseband TV-CIR can be written as
\begin{equation}\label{tvcir0}
  \tilde h(\tau,t)=\sum_{p=1}^{\tilde P}\tilde h_p(t) \delta (\tau-\tilde \tau_p(t)),
\end{equation}
where $\delta$ denotes the Kronecker delta function and $\tau$ is the delay domain variable.

During the channel's stationary time when the time-variation of $\tilde \tau_p(t)$ accounting for ``delay drift" can be neglected and the time variation of ${\mathsf a}_p(t)$ is caused by a Doppler spread $\tilde \nu_p$ \cite{Hlawatsch2011}, we have $\tilde \tau_p(t)=\tilde \tau_p$ and
$\tilde h_p(t)=\tilde h_p e^{j2\pi \tilde \nu_p t}$. Then, we can rewrite the received complex-valued baseband signal in (\ref{bbyt_general}) as
\begin{equation}\label{bbyt}
  y(t)=\sum_{p=1}^{\tilde P}\tilde h_p x(t-\tilde \tau_p)e^{j2\pi \tilde \nu_p t},
\end{equation}
and the baseband TV-CIR in (\ref{tvcir0}) as
\begin{equation}\label{tvcir}
  \tilde h(\tau,t)=\sum_{p=1}^{\tilde P}\tilde h_p e^{j2\pi \tilde \nu_p t} \delta (\tau-\tilde \tau_p).
\end{equation}
The corresponding DD domain representation of the LTV channel is
\begin{equation}\label{sf}
  \tilde h(\tau,\nu)=\sum_{p=1}^{\tilde P}\tilde h_p \delta (\tau-\tilde\tau_p)\delta (\nu-\tilde\nu_p),
\end{equation}
where $\nu$ is the Doppler domain variable.

During the coherence time, when the channel's time-variation caused by $\tilde \nu_p$ can be further neglected, (\ref{sf}) becomes the familiar LTI channel's impulse response
\begin{equation} \label{hlti_offgrid}
  \tilde h(\tau)=\sum_{p=1}^{\tilde P} \tilde h_p \delta (\tau-{\tilde \tau}_p),
\end{equation}
which only introduces time dispersion.
The IO relation of the LTI channel can be \emph{exactly} written as a \ac{1d} convolution between $x(t)$ and the time-invariant impulse response $\tilde h(\tau)$ in (\ref{hlti_offgrid}), given by
\begin{equation}
  y(t)=\sum_{p=1}^{\tilde P}\tilde h_p x(t-\tilde \tau_p)=\int_{-\infty}^{\infty} x(t-\tau)\tilde h(\tau) d\tau.
\end{equation}

Being a \ac{2d} discrete function in the DD domain, $\tilde h(\tau,\nu)$ in (\ref{sf}) represents a deterministic channel model and it is known as a special case of the spreading function $\mathcal S(\tau,\nu)$, which usually is a 2D continuous function characterizing a continuum of scatters\cite{Hlawatsch2011}. The statistical counterpart of the spreading function $\mathcal S(\tau,\nu)$ for the LTV channel is the scattering function given by \cite{bello}
\begin{equation}
  \mathcal C(\tau, \dot \tau; \nu, \dot \nu)=\mathbb E\left\{S(\tau,\nu)S^*(\dot \tau,\dot \nu)\right\}.
\end{equation}
When $\mathcal C(\tau,\dot \tau; \nu, \dot \nu)=\mathsf C(\tau,\nu)\delta (\tau-\dot \tau)\delta (\nu-\dot \nu)$,
we obtain the well-known wide-sense stationary uncorrelated scattering channel.
In this paper, we assume that communications occur in the stationary time interval with a deterministic channel model represented by the spreading function, instead of a random LTV channel model represented by an ensemble of spreading functions.

\begin{figure*}
  \centering
  \includegraphics[width=17.5cm]{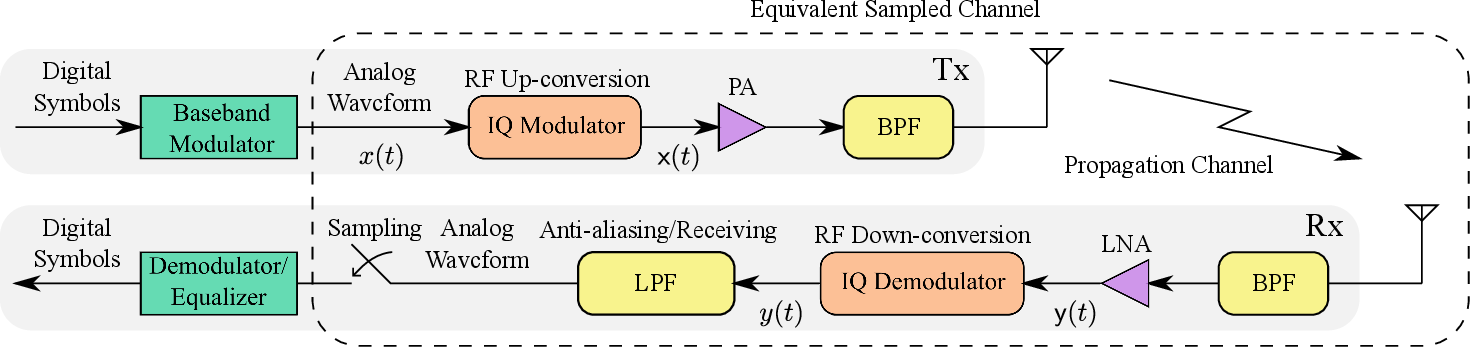}
  \caption{Block diagram of a wireless communication link}
  \label{eschannel}
\end{figure*}

Given a general spreading function $\mathcal S(\tau,\nu) $, the received complex-valued  baseband signal in (\ref{bbyt}) can be generalized to
\begin{equation}
  y(t)=\int_{-\infty}^{\infty} \int_{-\infty}^{\infty} \mathcal S(\tau,\nu) x(t-\tau)e^{j2\pi \nu t} d\tau d\nu,
\end{equation}
whose \ac{fd} representation is
\begin{equation}
  Y(f)=\int_{-\infty}^{\infty} \int_{-\infty}^{\infty} \mathcal S(\tau,\nu) X(f-\nu)e^{-j2\pi \tau(f-\nu)} d\tau d\nu.
\end{equation}
Meanwhile, given a normalized analysis window $g(t)$, the STFT of the transmitted signal $x(t)$ is defined as
\begin{equation}\label{stft}
  X^{(g)}(t,f) \triangleq \int_{-\infty}^{\infty} x(\dot t) g^*(\dot t-t)e^{-j2\pi f \dot t} d \dot t.
\end{equation}
Then, the STFT of the received signal $y(t)$ becomes \cite{Hlawatsch2011}
\begin{equation}\label{stft_io}
  Y^{(g)}(t,f)=\int\displaylimits_{-\infty}^{\infty} \int\displaylimits_{-\infty}^{\infty} \mathcal S(\tau,\nu) X^{(g)}(t-\tau, f-\nu)e^{-j2\pi \tau (f-\nu)} d\tau d\nu.
\end{equation}
It is interesting to observe that \emph{except for the phase term} $e^{-j2\pi \tau (f-\nu)}$, the above STFT-based IO relation of the LTV channel is a 2D convolution.

In mobile communications, the LTV channel is typically underspread with a spreading function confined to a small region in the DD domain. In particular, let $\tau_{\textrm{max}}$ and $\nu_{\textrm{max}}$ denote the channel's delay spread and Doppler spread, respectively.
Then, an LTV channel is said to be underspread, when $4\tau_{\textrm{max}}\nu_{\textrm{max}}\le 1$ \cite{Hlawatsch2011}.
Also, it should be noted that $\tilde \tau_p$ and $\tilde \nu_p$ are generally \emph{off-grid} and
the LTV channel $\mathcal S(\tau,\nu)$ or $\tilde h (\tau,\nu)$ is \emph{neither time- nor band-limited}, as evidenced by its TF representation\cite{Hlawatsch2011}
\begin{equation}
  \mathcal L_{\mathcal S}(t, f)=\int_{-\infty}^{\infty} \int_{-\infty}^{\infty} \mathcal S(\tau,\nu) e^{j2\pi(t\nu-f\tau)} d\tau d\nu.
\end{equation}

\subsection{Equivalent Sampled Channel Models}
Fig. \ref{eschannel} illustrates the block diagram of a typical wireless communication link. At the transmitter, even though the baseband signal $x(t)$ is designed to be band-limited, to mitigate the OOBE induced by the  non-linearity of \ac{pa}, the output of the PA is usually filtered by a \ac{bpf} to avoid leakage to adjacent channels and to ensure compliance with the spectrum masks.
At the receiver, the use of BPF is also a \emph{must} in order to
reject adjacent channel interference and out-of-band noise. In direct-conversion receivers, this operation is realized by the combination of a BPF and \ac{lpf} after RF down-conversion. For heterodyne receivers, a cascade of an image-rejection filter, mixer, and channel selection filter after the \ac{lna} is typically used\cite{rfme}. It is clear that regardless of the transceiver design, the band-limited nature of $x(t)$ and the filtering operations lead to a band-limited channel and system.

Bearing in mind that $x(t)$ is also time-limited, in the communication link of Fig. \ref{eschannel}, we can only observe an \mbox{\emph{effective}} channel that is the time- and band-limited version of the propagation channel. In other words, the channel that really matters is not the propagation channel but the effective channel, which is the cascade of the transmit filter, propagation channel, and receive filter. The concept of effective channel is straightforward for \ac{sc} modulation, since its transmit/receive filter is just the transmit/receive pulse. For MC modulation, the effective channel model is also valid, provided that the  transmit/receive filter represents the overall effect of the subcarrier-wise transmit/receive pulses/filter, for example, an ideal LPF having a passband bandwidth of $B_x$.

Note that baseband signal processing is typically conducted by \ac{dsp}, and therefore requires an appropriate sampling of the received basedband signal $y(t)$.
The frequency dispersion of the LTV channel will expand the bandwidth of $y(t)$ beyond $B_x$, but this is usually ignored in practice, since the Doppler spread is relatively small (on the order of tens to hundreds Hz) compared to the bandwidth $B_x$ (on the order of MHz)\cite{fwc}.

For SC modulation, the LPF can act as the receive \emph{matched filter}, the output of which is sampled at the symbol rate. The symbol rate (or sampling rate) is usually \emph{lower than $B_x$ to intentionally cause aliasing} and then achieve zero \ac{isi} over an ideal band-limited channel.
For MC modulations, a DSP-based low-complexity implementation is preferred, because the direct implementations of subcarrier-wise receive pulses/filters are expensive. In this case, usually a Nyquist sampling rate is adopted to strike a balance between the link performance and its implementation cost. Then, the LPF acts as an \emph{anti-aliasing filter}, which may be omitted if $y(t)$ has been appropriately filtered in the \ac{iq} demodulator.

It is clear now that although the LTV channel $\mathcal S(\tau,\nu)$ or $\tilde h (\tau,\nu)$ is of off-grid nature and neither time- nor band-limited, the combined duration and bandwidth constraints of $x(t)$ lead to \emph{a finite number of discrete-time samples} at the output of the effective channel, which results in an equivalent \emph{on-grid} channel, namely the ESDD channel, as shown in the dashed block of  Fig. \ref{eschannel}. {In other words, the DD domain considering a practical time- and band-limited signal is always a gridded DD domain associated with the specific delay (time) and Doppler (frequency) resolutions.}

Let the received signal's sampling rate and the time span of utilized samples be $\mathbb W$ and $\mathbb T$, respectively. The ESDD channel corresponding to the DD domain channel in (\ref{sf}) may be written as \cite{bello}
\begin{equation}\label{esddchannel}
  h(\tau,\nu)=\sum_{p=1}^{P}h_p \delta (\tau-\tau_p)\delta (\nu-\nu_p),
\end{equation}
where we have $P\ge \tilde P$, $\tau_p=l_p/\mathbb W$, $\nu_p=k_p/\mathbb T$, $l_p, k_p \in \mathbb Z$, and $1/\mathbb W$ as well as $1/\mathbb T$ are known as the delay and Doppler resolutions, respectively.
The TF representation of (\ref{esddchannel}) is given by
\begin{equation}
  \mathcal L_h(t, f)=\sum_{p=1}^{P} h_p e^{j2\pi(t\nu_p-f\tau_p)}
\end{equation}
for $t\in [0,\mathbb T]$ and $f\in [-\frac{\mathbb W}{2},\frac{\mathbb W}{2}]$.
Let us assume that the paths are arranged in ascending order of delay and $\tau_1=0$. Then the delay spread of the ESDD channel is given by $\tau_{\textrm{max}}=\tau_P$, and the LTI version of (\ref{esddchannel}) is given by
\begin{equation}\label{hlti_ongrid}
  h(\tau)=\sum_{p=1}^P h_p \delta (\tau-\tau_p),
\end{equation}
whose FD representation, i.e. the channel's transfer function is
\begin{equation}\label{Hlti_ongrid}
  H(f)=\sum_{p=1}^P h_p e^{-j2\pi \tau_pf},
\end{equation}
for $f\in [-\frac{\mathbb W}{2},\frac{\mathbb W}{2}]$. Note that the relation between (\ref{hlti_ongrid}) and (\ref{hlti_offgrid}) can be found in \cite{fwc} when the transmit filter is an ideal LPF having a passband bandwidth of $\mathbb W$.

The impact of the LTV and the LTI channels imposed on a \emph{practical} modulation waveform is characterized by the equivalent sampled channel models in (\ref{esddchannel}) and (\ref{hlti_ongrid}), respectively. Bearing in mind  these impacts, a modulation scheme entails an appropriate (bi)orthogonal pulse design to strike a comprise between the bandwidth efficiency and the complexity of demodulation and equalization.
In the following sections, we will discuss the modulation designs conceived for these channels.

\begin{table}
  \centering
  \caption{MC Modulation Parameters }
  \bgroup
  \def\arraystretch{1.3}
  \begin{tabular}{|c|r|}
    \hline
    \thead{Notation}                                                                                                                                                                                                    & \thead{Parameter}                                                                                                  \\
    \hline
    $\Delta F$                                                                                                                                                                                                          & \makecell[r]{frequency resolution,  subcarrier spacing,             \\ fundamental frequency (Ref. Figs. 1 and 4)} \\
    \hline
    $T$                                                                                                                                                                                                                 & \makecell[r]{symbol period, $T\equiv 1/\Delta F$ (Ref. Fig. 4)}                                                    \\
    \hline
    $\Delta T$                                                                                                                                                                                                          & \makecell[r]{time resolution, symbol interval (Ref. Figs. 1 and 4)}
    \\
    \hline
    $\Delta R$                                                                                                                                                                                                          & JTFR, $\Delta R=\Delta T\Delta F$                                                                                  \\
    \hline
    $N$                                                                                                                                                                                                                 & number of subcarriers                                                                                              \\
    \hline
    $M$                                                                                                                                                                                                                 & number of symbols                                                                                                  \\
    \hline
    $g(t)$                                                                                                                                                                                                              & \makecell[r]{transmit prototype pulse  (Ref. Fig. 3)}                                                              \\
    \hline
    $T_g$                                                                                                                                                                                                               & duration of $g(t)$, symbol duration (Ref. Figs. 3 and 4)                                                           \\
    \hline
    $G(f)$                                                                                                                                                                                                              & Fourier transform of $g(t)$ (Ref. Fig. 3)                                                                          \\
    \hline
    $B_g$                                                                                                                                                                                                               & bandwidth of $g(t)$, span of $G(f)$  (Ref. Figs. 3 and 4)                                                          \\
    \hline
    $B_x$                                                                                                                                                                                                               & bandwidth of $x(t)$   (Ref. Fig. 4)                                                                                \\
    \hline
    $T_x$                                                                                                                                                                                                               & duration of $x(t)$    (Ref. Fig. 4)                                                                                \\
    \hline
    $\gamma (t)$                                                                                                                                                                                                        & receive prototype pulse                                                                                            \\
    \hline
    $\mathbb W$\tablefootnote{$\mathbb W$ and $\mathbb T$ are also the nominal bandwidth and duration, respectively. They are different from $B_x$ and $T_x$, and adhere to $\mathbb W\le B_x$ and $\mathbb T\le T_x$.} & sampling rate at receiver (Ref. Section VI)                                                                        \\
    \hline
    $\mathbb T$                                                                                                                                                                                                         & time span of utilized samples at receiver (Ref. Section VI)                                                        \\
    \hline
  \end{tabular}
  \egroup
  \label{tab:parameters}
\end{table}

\section{TFMC Modulation Schemes Designed for LTI Channels}
Starting from 1950s, MC modulation techniques including the OFDM have been developed for more than half a century.
In the literature, there are many technical reviews of OFDM and its applications, such as \cite{ofdm,tff,mct,keller_proc_ieee_2000,jiang_proc_ieee_2007,weinstein2009,mc_book_lly_2009}. The interested reader can find a comprehensive historic evolution of OFDM in \cite[Table II]{jiang_proc_ieee_2007}.

The main purpose of this paper is to study the new ODDM/DDMC modulation, the corresponding new pulse design, and the unique transmission strategy for LTV channels.
It is known that the (bi)orthogonal pulse design in conventional TFMC modulation schemes is closely related to the properties of a (bi)orthogonal WH function set, which is governed by the WH frame theory\cite{tff}. In this section, we review the conventional TFMC modulation schemes, in particular, from the perspective of the eigenfunction-based transmission strategy, the WH frame theory based (bi)orthogonal pulse design principles, and the implementation methods.

\subsection{Pulse Design Principles}
The transmit pulses in an MC modulation can be represented by a function set\cite{tff,primer}
\begin{equation}\label{gtf}
  \left(g,\Delta T, \Delta F\right)=\left\{g_{m,n}\right\}_{m,n\in \mathbb Z},
\end{equation}
where $g_{m,n}\triangleq g(t-m\Delta T)e^{j2\pi n \Delta F (t-m\Delta T)}$, $g(t)$ is the prototype pulse, $\Delta T$ and $\Delta F$ are the time resolution (symbol interval) and frequency resolution (subcarrier spacing), respectively.
Similarly, we can form the receive pulses $\left(\gamma,\Delta T, \Delta F\right)$ using another prototype pulse $\gamma(t)$ having the same time and frequency resolutions. Meanwhile, the inverse of the frequency resolution is known as the \emph{symbol period} denoted by $T=1/\Delta F$.

Then, the transmit waveform of MC modulation synthesized by the transmit pulses in (\ref{gtf}) is given by
\begin{align}\label{xt}
  x(t)=\sum_{m=0}^{M-1}\sum_{n=-N/2}^{N/2-1} X[m,n]g(t-m\Delta T)e^{j2\pi n \Delta F (t-m\Delta T)},
\end{align}
where $M$ is the number of MC symbols contained by $x(t)$, and the number of subcarriers $N$ is usually supposed to be an even number. Furthermore,
$X[m,n]$ for $-N/2\le n \le N/2-1, 0\le m \le M-1$ represent the information-bearing digital symbols drawn from a signal constellation diagram, for example, \ac{qam}. Note that for MC systems, the modulator and demodulator are essentially inverse \ac{stft} and \ac{stft} in (\ref{stft}) and (\ref{stft_io}). Also, one can see from (\ref{xt}) that \ac{sc} modulation is a special case of MC modulation associated with $n\equiv 0$.

\begin{figure}
  \centering
  \includegraphics[width=8.8cm]{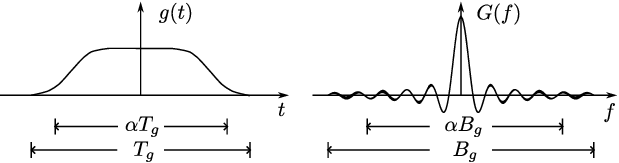}
  \caption{Metrics of pulse's TF occupancy}
  \label{dtdf}
\end{figure}

Let $T_g$ and $G(f)$ denote the duration and the Fourier transform of $g(t)$, respectively.
The main parameters of MC modulation are listed in Table \ref{tab:parameters}, where
$B_g$, the bandwidth of $g(t)$ or the span of $G(f)$, is also defined in the essential sense of \cite{onbandwidth} to capture the majority of the energy of the pulse. However, from a strictly mathematical point of view, a pulse cannot simultaneously have limited duration $T_g$ and limited bandwidth $B_g$.
Meanwhile, it is noteworthy that apart from $T_g$ and $B_g$, a classic metric of the occupancy of $g(t)$ in the TF domain is its \ac{tfa} $A_g =\alpha T_g\alpha B_g$, where the pulse's effective duration $\alpha T_g$ and effective bandwidth $\alpha B_g$ are defined as the standard deviations of its TD and FD shapes, respectively \cite{gabor,haas_wpc_97}.
As an example, the differences between $T_g$ and $\alpha T_g$,  $B_g$ and $\alpha B_g$ are shown in Fig. \ref{dtdf}.
Due to the Heisenberg uncertainty principle, the TFA obeys a lower bound $A_g\ge 1/(4\pi)$ known as the \emph{Gabor limit}, which is attained by the Gaussian pulse \cite{gabor}\footnote{Suffice to say that the Gaussian pulse exhibits the smallest occupancy in the TF domain. As a result, it was adopted by \ac{2g} \ac{gsm} communications across about 150 countries.}.
Usually, $g(t)$ is said to be well-localized in the sense of minimum TF energy spread, when its signal energy is concentrated around its centre to have a small TFA \cite{haas_wpc_97}.

Given $\Delta T$ and $\Delta F$, the fundamental issue of MC modulation is to find $g(t)$ and $\gamma(t)$ that whould satisfy the orthogonal condition of
\begin{equation}\label{g_ortho}
  \langle g_{m,n}, g_{\dot m,\dot n}\rangle =\delta(m-\dot m)\delta(n-\dot n),
\end{equation}
or the biorthogonal condition of
\begin{equation}\label{g_biortho}
  \langle g_{m,n}, \gamma_{\dot m,\dot n}\rangle =\delta(m-\dot m)\delta(n-\dot n).
\end{equation}
When (\ref{g_ortho}) or (\ref{g_biortho}) holds, the prototype pulses $g(t)$ and $\gamma(t)$ are said to be \emph{orthogonal or biorthogonal with respect to} the time resolution $\Delta T$ and the frequency resolution $\Delta F$.

By considering the TF domain as a 2D phase space, the function set in (\ref{gtf}) forms a discrete grid ``sampling" the phase space\cite{haas_wpc_97,wavelet}, where the sampling resolution is the JTFR $\Delta R=\Delta T\Delta F$.
Then, the function set in (\ref{gtf}) is treated as a WH set, the density of which is given by
the inverse of the JTFR as $\mathcal D=\Delta R^{-1}$.
From the WH frame theory, the existence of (bi)orthogonal WH sets depends on the sampling resolution $\Delta R$, which can be summarized as \cite{cofdm,wavelet,ftfa,gaborana,wexler1990,haas_wpc_97,kozek98,tff,strohmer2003}:
\begin{itemize}
  \item Critical sampling ($\Delta R =1$) : Orthogonal WH sets exist.  However, they have either infinite TD or FD energy spread according to the Balian-Low theory \cite{wavelet_book}. Therefore, they are not well-localized in the TF domain.
  \item Under-critical sampling ($\Delta R >1$) : Orthogonal or biorthogonal WH sets exist, if $\Delta R$ is larger than $1$ to employ a TF guard region\cite{tff}.
  \item Over-critical sampling ($\Delta R <1$) : Neither orthogonal nor biorthogonal WH set exists.
\end{itemize}

Here, we define the bandwidth efficiency of the transmit signal $x(t)$ as
\begin{align}\label{se}
  \eta=\frac{MN}{B_xT_x}.
\end{align}
Since the functions $g_{m,n}$ in (\ref{xt}) are only different in terms of their TF centre,
it is clear that $\eta$ depends not only on the density of $\left(g,\Delta T, \Delta F\right)$ but also on the duration $T_g$ of $g(t)$, and on the bandwidth $B_g$ of $g(t)$.
In other words, to achieve high bandwidth efficiency, we have to place the pulses
as densely as possible while keeping them (bi)orthogonal, which subsequently requires a fine JTFR and a well-localized $g(t)$. In fact, the highest bandwidth efficiency corresponds to the best use of the available dimension, which is also known as the \ac{dof} of time- and band-limited signals\cite{Jacobs_pce,fwc,onbandwidth}.

\subsection{Eigenfunction-based Transmission over LTI Channels}

Passing the MC signal $x(t)$ of (\ref{xt}) through the LTI channel of (\ref{hlti_ongrid}), the received waveform is given by
\begin{align}
  y(t)=\int_{-\infty}^{\infty}x(t-\tau)h(\tau) d\tau=\sum_{p=1}^P h_p x(t-\tau_p).
\end{align}
Then, the receive pulse $\gamma_{m,n}$ is applied to $y(t)$ for extracting the signal component at the $(m,n)$-th TF grid point
\begin{align}\label{rp}
  Y[m,n]=\int_{-\infty}^{\infty}y(t)\gamma^*(t-m\Delta T)e^{-j2\pi n \Delta F (t-m\Delta T)} dt,
\end{align}
which are fed into the channel equalizer of Fig. \ref{eschannel} to recover the transmitted digital symbols.

For SC modulation associated with $n=0$, the transmit pulses occupy the whole bandwidth. Hence, the channel equalization is generally expensive, especially for channels exhibiting severe frequency selectivity.
On the other hand, by slicing a wideband frequency selective channel into multiple narrowband frequency-flat subchannels, simple single-tap FD equalizers may be used, as in OFDM\cite{ofdm}.

The rationale behind the single-tap equalization of OFDM is due to its eigenfunction-based transmission strategy\cite{fwc}.
Recall that the subcarriers in OFDM or in general MC modulations are complex sinusoids, which are the eigenfunctions of LTI systems. Their frequencies are deliberately selected to be \emph{integer multiples} of the frequency resolution $\Delta F$, which leads to the term $n\Delta F$ in (\ref{xt}). {Because of the discrete upper harmonic frequencies of the subcarriers, OFDM is also known as \ac{dmt} modulation, especially in the \ac{dsl} technology\cite{adsl}.} Meanwhile, observe that the symbol period $T=1/\Delta F$ is the common or fundamental period of these sinusoids,
the orthogonal pulses can be obtained via truncating the subcarriers or equivalently modulating the subcarriers with the prototype pulse $g(t)=\rect_T(t)$ \cite{toiotf_harmuth_60,ofdm_dft_weinstein_71,mc_commag_90}. Clearly, these truncated subcarriers are orthogonal to each other within the symbol period $T$, and each MC symbol is basically a single cycle of a periodic signal having a period of $T$.

Considering the time dispersion of the channel, the symbol duration $T_g$, which is also the length of the prototype pulse, is extended from $T$ to $T+\tau_{\textrm{max}}$.
The pulses' extra length corresponding to the channel's  delay spread $\tau_{\textrm{max}}$ is a necessary \emph{TD redundancy}, which prevents ISI induced by the time dispersion of the channel and then guarantees the orthogonality across \emph{a symbol period, namely the effective part} of these truncated eigenfunctions.
Since each MC symbol is a part of a periodic signal, the extension of the transmit pulses is equivalent to prepending a $T_{\textrm{cp}}$-length \ac{cp} \cite{cp}, corresponding to $g(t)=\rect_{T_g}(t)$ where $ T_g=T_{\textrm{cp}}+T$ and $T_{\textrm{cp}}\ge \tau_{\textrm{max}}$. This type of OFDM is called CP-OFDM or just OFDM for simplicity. Furthermore, since the CP-OFDM symbols are sent one after another, we have $\Delta T=T_g$.
Given these parameter settings, each CP-OFDM symbol  \emph{emulates} an eigenfunction-based input that results in a scalar channel IO relation.
In particular, given the receive prototype pulse $\gamma(t)=\rect_T(t-T_{\textrm{cp}})$, we have $Y[m,n]\approx H(n\Delta F)X[m,n]$, which is free of ISI and of \ac{ici}, enabling an easy recovery of $X[m,n]$.

\begin{figure}
  \centering
  \includegraphics[width=6.8cm]{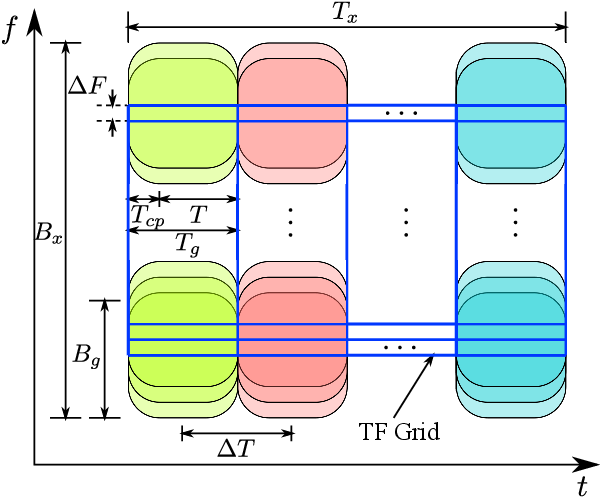}
  \caption{TF grid and signal localization of CP-OFDM}
  \label{tf_lattice}
\end{figure}

Let a $B_g\times T_g$ block represent the occupancy of $g(t)$ in the TF domain. The corresponding TF grid and the TF signal localization of CP-OFDM with $g(t)=\rect_{\Delta T}(t)$ is illustrated in Fig. \ref{tf_lattice}, where we obtain $T_x =MT_g= M\Delta T$ and $B=(N-1)\Delta F+B_g > N\Delta F$, because $B_g>\Delta F$. Given a fixed $T_x$, {the TD redundancy causes the reduction of the number of symbols} and consequently erodes the bandwidth efficiency. Meanwhile, without explicitly presenting the bandwidth and duration of each pulse, the TF grid is usually treated as a \emph{simplified} version of the TF signal localization\cite{tff}, as shown earlier in Fig. \ref{tf_grid}.

\subsection{Classification of Pulses and TFMC Schemes}\label{unload_subcarrier}

Given that the pulse density obeys $\mathcal D=\Delta R^{-1}$, it is clear that $g(t)$ satisfying the orthogonal condition of (\ref{g_ortho}) for the critical sampling of $\Delta R=1$ achieves the highest pulse density.
A plausible choice of such a $g(t)$ is the aforementioned rectangular pulse $\rect_T(t)$ first proposed in \cite{toiotf_harmuth_60}, which corresponds to the CP-free OFDM. Since its duration is constrained to the symbol period $T$,  $\rect_T(t)$ belongs to the \emph{\ac{tlop}} set\cite{tlomcm_95}, where $T_g$ is usually less than $2T$. {However, $\rect_T(t)$ theoretically has its energy spreading to infinity in the FD caused by its slowly decaying $\sinc$ shaped spectrum}, which agrees with the Balian-Low theory \cite{wavelet_book}.
In practical OFDM systems having stringent spectral restrictions, some edge subcarriers are unloaded to suppress the OOBE and to ease the transmit and receive filtering \cite{80211a,80211ac,3gpp_4g}.
Moreover, we may have to further sharpen the signal spectrum by applying a frequency localized TD window after cyclically extending the OFDM symbol by both a CP and a \ac{cs} \cite{mallory_92,tlomcm_95,80211a,80211ac,rcwofdm,wola,waveformcandidate}. Upon further taking the necessary TD redundancy into account, the cyclic extension of each OFDM symbol results in a lower time resolution namely an extended symbol interval of $\Delta T> T=1/\Delta F$, implying the spectrally inefficient under-critical sampling of $\Delta R >1$. As a result, by considering the vacant  subcarriers in the FD and the cyclic extension in the TD, the TLOP-based OFDM suffers from a considerable loss of bandwidth efficiency in practice, due to the lack of well-localized orthogonal pulse in the case of critical sampling.

The somewhat disappointing spectral containment of the TLOP has motivated the design of a \emph{\ac{blop}} \cite{ofdm_oqam_chang_66,ofdm_oqam_saltzberg_67,ofdm_oqam_hirosaki_81,ofdm_oqam_hirosaki_86,finitepulse_96,oqam_pulse_icc99,ofdm_oqam_siohan_02,fbmcprimer}, where the constraint imposed on the pulse's duration $T_g$ is relaxed from $2T$ to \emph{multiple symbol periods},
for better approximating their theoretically infinite duration. 
Since the pulse's duration $T_g$ becomes much longer than the symbol interval $\Delta T$, a heavy overlap of pulses occurs in the TD. As a result, the BLOP-based OFDM is also known as \ac{smt} modulation\cite{ofdmvsfbmc,mct}. In BLOP-based OFDM, in exchange for the TD overlapping, only the neighboring FD subchannels are overlapped with each other \cite{ofdm_oqam_chang_66}. This is in contrast  to the densely overlapped subchannels of TLOP-based OFDM shown in Fig. \ref{tf_lattice}.
Also, in contrast to the classic QAM signaling in TLOP-based OFDM, BLOP-based OFDM usually employs \ac{oqam} signaling to retain the orthogonality with well-localized pulses, and is termed as OFDM/OQAM\cite{weinstein2009}. Furthermore, if we treat \emph{$g(t)$ as the impulse response of a filter}, the frequency-shifted pulses actually form a filter bank. Therefore, the OFDM/OQAM is also known as \ac{fbmc} with OQAM (FBMC/OQAM). Similarly, the TLOP-based OFDM associated with a well-localized $g(t)$ rather than with the ordinary rectangular one is called FBMC or \ac{psofdm} \cite{zhao_pulse_2017}.

With a well-localized pulse $\mathpzc g(t)$, OFDM/OQAM achieves excellent FD containment in the case of critical sampling.
The ``little magic" of the OFDM/OQAM in terms of circumventing the Balian-Low theory \cite{oqam_pulse_icc99} involves shortening
the symbol interval to $T/2$, and at the same time replacing the \emph{complex-valued} orthogonality in (\ref{g_ortho}) with \emph{real-valued} orthogonality, which can be written as \cite{ofdm_oqam_siohan_02}
\begin{equation}\label{r_ortho}
  \langle \mathpzc  g_{m,n}, \mathpzc g_{\dot m,\dot n}\rangle_{\Re} \triangleq \Re\{\langle \mathpzc  g_{m,n}, \mathpzc g_{\dot m,\dot n}\rangle\}=\delta_{m-\dot m,n-\dot n},
\end{equation}
where $\mathpzc g_{m,n}\coloneqq \mathpzc g(t-m\frac{T}{2})e^{j2\pi n \Delta F (t-m\frac{T}{2})}e^{j\phi_{m,n}}$, and $\phi_{m,n}$ is a phase term to ensure that the interference becomes purely imaginary under OQAM signaling.
In addition, due to the pulse's long duration, the channel-induced ISI is usually negligible, and therefore the CP used for ISI mitigation can be omitted. As consequence, OFDM/OQAM has the highest bandwidth efficiency and still exhibits robustness against the channel's time dispersion.

In summary, TFMC modulations designed for LTI channels adopt the eigenfunction-based transmission strategy to ease the equalization of Fig. \ref{eschannel}. According to the different eigenfunction-based pulse designs, they may be categorized as the TLOP-based OFDM (OFDM or CP-OFDM, FBMC or PS-OFDM) and the BLOP-based OFDM (OFDM/OQAM, FBMC/OQAM or SMT).
A comparison of these popular TFMC modulation schemes is shown in Table \ref{tab:mcmclassification}. Hereafter, we will use the terms of OFDM, PS-OFDM and OFDM/OQAM, for the sake of simplicity.

It is noteworthy that in contrast to the eigenfunction-based transmission of TFMC modulation, the conventional SC modulation is an impulse-function-based transmission. In particular, its transmit pulses are basically the variants of \emph{band-limited impulses}, and lead to $T_g>\Delta T$. Usually, these transmit pulses will lose mutual orthogonality in the presence of the channel's time dispersion, and result in a \emph{convolutional} system IO relation. Meanwhile, it has been known that the SC modulation can also be combined with TD redundancy to achieve simple FD equalization\cite{scfde} and flexible multi-user band allocation\cite{Sari_SC_1995,goodman_sc_fdma_2006}. Given its benefits, it was adopted in the 4G standard\cite{4Gbook}.

Now one can see that for LTI channels, generally we have two choices of transmission strategy: either the eigenfunction-based one having reduced bandwidth efficiency and simple channel equalization, or the impulse-function-based one having higher bandwidth efficiency but relatively complex channel equalization.
In other words, there is always a trade-off between the bandwidth efficiency and the equalization complexity.

\begin{table}
  \def\arraystretch{3.3}
  \centering
  \caption{Comparison of Popular TFMC Modulation Schemes}
  \begingroup
  \setlength{\tabcolsep}{0.6em}
  \begin{tabular}{|c|c|c|c|c|}
    \hline
    \thead{Scheme}                                                                                                  & \thead{Pulse}                                                              & \thead{\makecell{Cyclic extension, \\ pulse duration, \\ symbol interval }} & \thead{Signaling} \\
    \hline
    \makecell{OFDM                                                      \\ (CP-OFDM), \\Filtered OFDM\footnotemark} & \makecell{Ordinary rectangular  \\ pulse with vacant \\ edge subcarriers } & \makecell{ CP, \\ $T<T_g<2T$, \\ $\Delta T=T_g$ }                           & QAM               \\
    \hline
    \makecell{PS-OFDM,                                                  \\FBMC}                                     & \makecell{TF well-localized \\ TLOP}                                       & \makecell{CP+CS, \\$T<T_g<2T$, \\ $T<\Delta T \le T_g$}                     & QAM               \\
    \hline
    \makecell{OFDM/OQAM,                                                \\FBMC/OQAM, \\SMT }                        & \makecell{TF well-localized \\ BLOP}                                       & \makecell{No CP,  \\$T_g>2T$, \\ $\Delta T=T/2$}                            & OQAM              \\
    \hline
  \end{tabular}
  \endgroup
  \label{tab:mcmclassification}
\end{table}
\footnotetext{This is explained in  Section \ref{tfmc_imple_method}.}

\subsection{Implementation Methods}\label{tfmc_imple_method}
Given $\Delta T$, $\Delta F$ and $g(t)$, how to generate the waveform in (\ref{xt}) for MC modulation at low complexity is of pivotal practical importance, especially when the number of subcarriers $N$ is large. In theory, we have two direct \emph{analog} approaches using $N$ modulators associated with carrier frequencies of $n\Delta F, n=-\frac{N}{2},\cdots \frac{N}{2}-1$. As shown in Fig. \ref{adi1}, one of them is to generate $X[m,n]g(t-m\Delta T)$, modulate it according to  $e^{j2\pi n \Delta F (t-m\Delta T)}$, and then add them up to obtain
\begin{align}\label{xmt}
  x_m(t)=\sum_{n=-N/2}^{N/2-1} X[m,n]g(t-m\Delta T)e^{j2\pi n \Delta F (t-m\Delta T)},
\end{align}
where $g(t)$ is treated as a \emph{transmit pulse/filter}.
The other one shown in Fig. \ref{adi2} is to generate $X[m,n]e^{j2\pi n \Delta F (t-m\Delta T)}$, add them up to have
\begin{align}\label{txmt}
  \tilde x_m(t)= \sum_{n=-N/2}^{N/2-1}X[m,n]e^{j2\pi n \Delta F (t-m\Delta T)},
\end{align}
and then truncate the result by the pulse $g(t-m\Delta T)$ to obtain $x_m(t)=\tilde x_m(t) g(t-m\Delta T)$, where $g(t)$ is treated as a \emph{prototype pulse or window function}.
Once $x_m(t), m=0,\cdots,M-1,$ are available, we can send them to a \ac{tmx} parameterized by the symbol interval $\Delta T$ to obtain $x(t)=\sum_{m=0}^{M-1}x_m(t)$, as in (\ref{xt}). The filter bank of Fig. \ref{adi1} and the pulse-shaping seen in Fig. \ref{adi2} correspond to the terminology of FBMC and PS-OFDM, respectively.

\begin{figure}
  \centering
  \includegraphics[width=8.5cm]{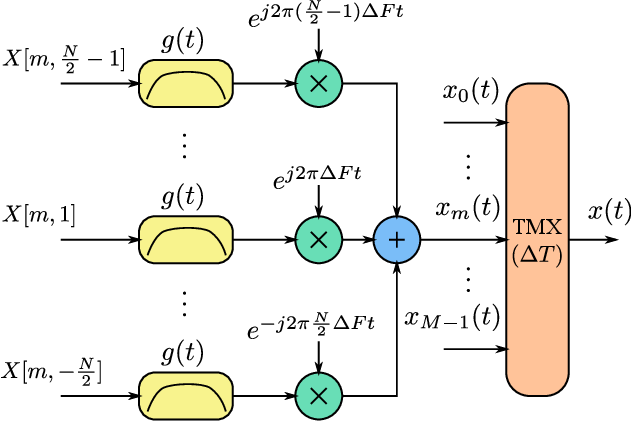}
  \caption{Analog implementation with $g(t)$ as transmit pulse/filter}
  \label{adi1}
\end{figure}

\begin{figure}
  \centering
  \includegraphics[width=8.5cm]{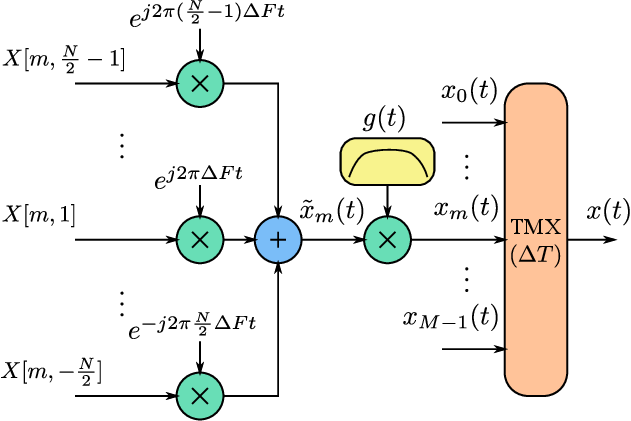}
  \caption{Analog implementation with $g(t)$ as prototype pulse/winow function}
  \label{adi2}
\end{figure}

However, these two methods require $N$ modulators, and therefore have prohibitively high implementation complexity. The implementation of MC modulation was considered unreasonable, until the connection between the \ac{idft}, the \ac{dft}, and the MC modulation/demodulation was found in \cite{zimmerman_kathryn_1967}, where an analog hardware based implementation of the IDFT was employed to generate the MC waveform.
After that, the \ac{ifft} and \ac{fft} algorithms \cite{fft} were introduced in \cite{salz1969} to realize the IDFT and DFT, which leads to the widely adopted IFFT/FFT-based implementation of MC modulation \cite{ofdm_dft_weinstein_71}.

\begin{figure}[t]
  \centering
  \includegraphics[width=8.5cm]{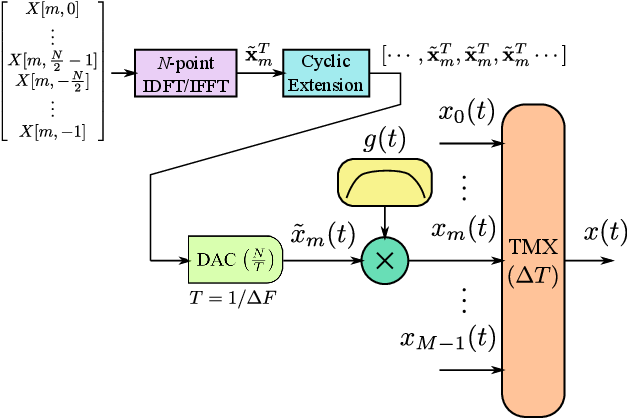}
  \caption{Digital implementation with $g(t)$ as prototype pulse/window function}
  \label{adi2n_ifft1}
\end{figure}

As shown in Fig. \ref{adi2n_ifft1}, the generation of $x(t)$ or $x_m(t)$ using the IFFT relies on a digital implementation of the second analog approach of Fig. \ref{adi2}, which is based on the following two observations. The first observation is that although $x_m(t)$ has a bandwidth of $B>N\Delta F$, $\tilde x_m(t)$ in (\ref{txmt}) is strictly band-limited to $[-\frac{N}{2}\Delta F, (\frac{N}{2}-1)\Delta F]$. Because the highest frequency is $\frac{N}{2}\Delta F$, sampling $\tilde x_m(t)$ at the Nyquist rate of $N\Delta F=N/T$ becomes \emph{feasible}, {by obeying the sampling theorem.} Then, the $N$ samples of $\tilde x_m(t)$ within one period $T=\frac{1}{\Delta F}$ are given by
\begin{align}
  \tilde x_m[\dot n] \triangleq  \tilde x_m \left(m\Delta T+\dot n \frac{T}{N}\right) = \sum_{n=-N/2}^{N/2-1}X[m,n]e^{j2\pi \frac{n \dot n}{N}}
\end{align}
for $0\le \dot n \le N-1$, which exactly represent the IDFT of
$[X[m,0],\cdots,X[m,\frac{N}{2}-1],X[m,-\frac{N}{2}],\cdots,$ $X[m,-1]]^T$.
The second observation is that $\tilde x_m(t)$ is an \emph{infinite-length periodic} signal, which indicates that we can repeat $N$ samples in one period to obtain the samples of $\tilde x_m(t)$.
Let $\tilde {\mathbf x}_m= [\tilde x_m[0], \cdots, \tilde x_m[N-1]]^T$. Then $\tilde x_m(t)$ can be generated by passing the cyclic extension of $\tilde {\mathbf x}_m^T$ namely $[\cdots, \tilde {\mathbf x}_m^T, \tilde {\mathbf x}_m^T, \tilde {\mathbf x}_m^T \cdots]$ through an ideal LPF with passband bandwidth $\frac{N}{T}$ \cite{oddmicc22}, which is actually the interpolation filter in the \ac{dac} having a rate of $\frac{N}{T}$. After that, the $g(t)$-based windowing is applied to $\tilde {\mathbf x}_m(t)$ to obtain $x_m(t)$, whose bandwidth is then expanded to
$B=(N-1)\Delta F+B_g>N\Delta F$ when $B_g>\Delta F$. It should be noted that the windowing and time multiplexing can also be conducted in the digital domain before the DAC, if their implementation is performed at an \emph{oversampling} rate of $\mathsf K \frac{N}{T}$, where $\mathsf K>1$ is an integer representing the oversampling factor. The oversampling based implementation of MC modulation is shown in Fig. \ref{adi2n_ifft2}.

\begin{figure}[t]
  \centering
  \includegraphics[width=8.5cm]{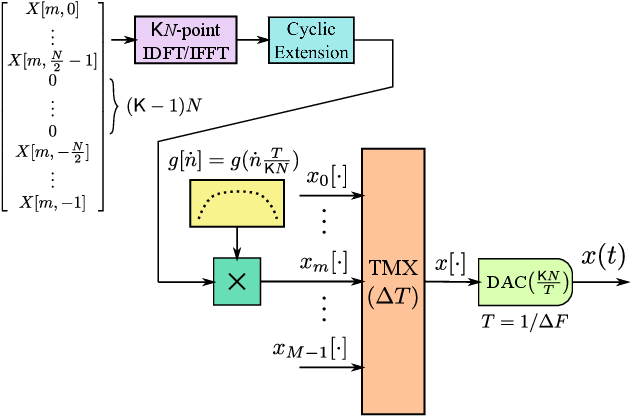}
  \caption{Oversampling based digital implementation}
  \label{adi2n_ifft2}
\end{figure}

In practice, \emph{given the appropriate parameter settings}, the cyclic extension of $\tilde {\mathbf x}_m^T$ and the subsequent windowing can be omitted for further simplifying the implementation. Recall that $B=(N-1)\Delta F+B_g>N\Delta F$ and $N$ is usually a power of $2$ for the IFFT. Then $x_m(t)$ becomes \emph{approximately} band-limited to $[-\frac{N}{2}\Delta F, (\frac{N}{2}-1)\Delta F]$, if we have some unloaded  subcarriers at the band edge. In fact, a practical OFDM system often only has $\bar N < N$ subcarriers, resulting in
\begin{align}\label{txt_bl}
  x_m(t)= \sum_{\substack{n=-\bar N/2, \\ n \ne 0}}^{\bar N/2}X[m,n]g(t-m\Delta T)e^{j2\pi n \Delta F(t-m\Delta T)},
\end{align}
where the \ac{dc} subcarrier is also usually left unloaded to ease the RF circuit design\cite{rfme}. With $\dot N=N-\bar N$ vacant subcarriers at the band edge, the bandwidth of $x_m(t)$ is $B=\bar N\Delta F+B_g \le  N\Delta F$, where the setting of $B<N\Delta F$ actually corresponds to the aforementioned oversampling-based implementation of Fig. \ref{adi2n_ifft2}. Then, we can pass $\tilde {\mathbf x}_m^T$ (after prepending a CP for channel delay spread) through the ideal LPF to obtain $x_m(t)$ directly, as long as $g(t)$ is $\Pi_{\Delta T}(t)$, implying that the simple truncation is not needed for $\tilde {\mathbf x}_m^T$ which is already time-limited. This vacant subcarriers based implementation, {which has been widely adopted in practice because of its bandwidth flexibility}, is shown in Fig. \ref{adi2n_ifft3}.

\begin{figure}[t]
  \centering
  \includegraphics[width=8.5cm]{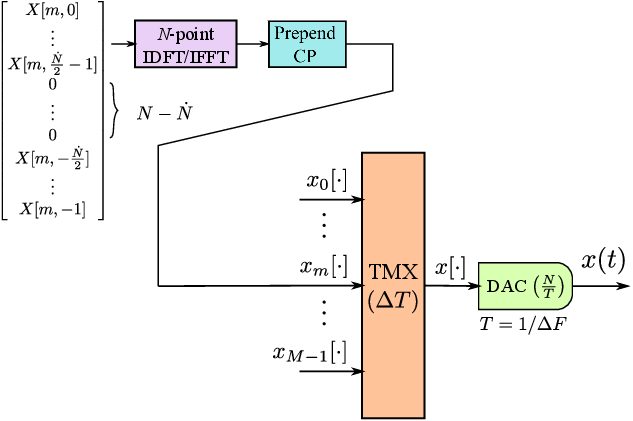}
  \caption{Vacant edge subcarriers based digital implementation (Filtered OFDM)}
  \label{adi2n_ifft3}
\end{figure}

Filtering or equivalently sample-wise pulse-shaping $\tilde {\mathbf x}_m^T$ by an LPF results in filtered OFDM\cite{fofdm,fofdmspawc}, where the orthogonality among subcarriers will be eroded, especially for those at the band edge. This problem can be avoided by leaving edge subcarriers blank\cite{fofdm}. As a result, filtered OFDM can be considered as OFDM associated with rectangular pulse and vacant edge subcarriers, as shown in Table \ref{tab:mcmclassification}.

It is also noteworthy that since the interpolation filter in a DAC is exactly an LPF, practical OFDM systems with the vacant subcarriers based implementation are actually filtered OFDM systems. The simulations of such OFDM systems are usually performed based on TD samples with $\frac{1}{N\Delta F}$-interval, which is appropriate only when there is a sufficiently large number of vacant edge subcarriers in the system.

In comparison to the convenient  IFFT-based implementation of the OFDM and PS-OFDM, the implementation of OFDM/OQAM has a considerably higher complexity, owing to the overlapping of MC symbols shaped by the long-duration $\mathpzc g(t)$.
An elegant approach to alleviate this difficulty is to amalgamate the IFFT with
a polyphase network for  efficiently realizing a filter bank \cite{ofdm_oqam_siohan_02,fbmcprimer}. This approach is  essentially a FD equivalent of the aforementioned TD operations, including the cyclic extension and the windowing/pulse-shaping.

\section{TFMC Modulation Schemes for LTV Channels}\label{tfmc_for_ltv}
For doubly-selective LTV channels, unfortunately, neither OFDM nor OFDM/OQAM performs well.
However, considering their wide deployment in practice, numerous efforts have been devoted to studying OFDM for transmission over doubly-selective channels.
A possible solution is to increase the subcarrier spacing, or equivalently, reduce the symbol period of the OFDM symbol. However, this approach may degrade the transmission efficiency due to the increased CP overhead when the CP length is fixed.

Moreover, based on different approximations of the TV-CIR and the resultant nondiagonal but banded FD channel matrix, the LTV channel {estimation} and equalization techniques can be rather complex (see \cite{cimini_tcom_1985,li_tcom_1998,ici_ofdm_tcom_1999,ced_mc_tcom2001,ici_ofdm_tcom_2003,lc_eq_ofdm_tsp_2004,1291798,1315913,1381441,1413242,turbo_eq_ofdm_tsp_2008, ov_ofdm_tcom_2011} and the references therein). It is pointed out that the off-diagonal elements of the FD channel matrix actually  offer \emph{time diversity {or Doppler diversity}} \cite{ced_mc_tcom2001,ici_ofdm_tcom_2003}.
However, these complex TV-CIR channel models often suffer from expensive yet somewhat inaccurate channel estimation. This is because the TV-CIR has a large number of parameters even along with a low-rank approximation.
For OFDM/OQAM, due to its intrinsic imaginary interference caused by the real-valued orthogonality constraint, the channel-induced ISI and ICI remain cumbersome.
To the best of our knowledge, there is no low-complexity equalization solution for OFDM/OQAM operating in high-mobility environments, where single-tap equalizers fail to achieve satisfactory performance \cite{nissel2017_cl}.

On the other hand, in the spirit of eigenfunction-based transmission, {PS-OFDM schemes for achieving} a scalar IO relation for transmission over LTV channels
can be found in \cite{haas_wpc_97,kozek98,strohmer2003,LTVDesign_2,matz07,jung07} and the references therein.
It is widely recognized that the underspead LTV channels at best have a structured set of approximate eigenfunctions \cite{eigen}.
Owing to the necessity to consider both time and frequency dispersions, these approximate eigenfunctions not only depend on the channel's spreading or scattering functions, but also require a well-localized prototype pulse $g(t)$ and a  much coarser JTFR than that of OFDM \cite{eigen,matz07}. In fact, the corresponding pulses can be considered as an approximation of the (bi)orthogonal WH sets in the case of under-critical sampling of $\Delta R>1$, and therefore require not only TD but also \emph{FD redundancy}.
Moreover, the approximate nature of these eigenfunctions leads to residual ISI and ICI, which may still remain cumbersome and have to be equalized carefully.
As a result, there are two main challenges in realizing eigenfunction-based transmission in LTV channels. The first one is the adaptation of transmit pulses to the channel, which is usually not possible in practical transmitters. The second one is the considerably reduced bandwidth efficiency if coarse JTFR is adopted to preserve eigen property.

In summary, the conventional TFMC modulations designed for eigenfunction-based transmissions in LTI channels suffer not only from costly and inaccurate channel estimation, but also from complex equalization in LTV channels. On the other hand, PS-OFDM designed for eigenfunction-based transmission in LTV channels have severe practical challenges and exhibit low bandwidth efficiency.

\section{MC Modulations for ESDD channels}
The challenges imposed by  eigenfunction-based transmission in LTV channels motivate us to consider new transmission strategies and develop alternative modulation schemes. To this end, it is necessary to reconsider the properties of LTV channels and design specifically ``tailored" channel-oriented pulses.

\subsection{Common Properties of ESDD Channels}
In practical transceivers, the transmit and receive pulses are usually fixed. Therefore, these pulses should be designed according to the \emph{common} properties of LTV channels.
However, in contrast to LTI channels relying on the complex sinusoidal eigenfunctions as their common properties, there is basically no common property for LTV or DD channels, due to the different propagation environments.

Recall that due to the limited bandwidth and duration of the signal, we only observe an ESDD channel at the receiver. Thus the ESDD channel is the one that matters for signal transmissions. Then, what we really care about is the common properties of ESDD channels,  rather than those of DD channels.
As shown in (\ref{esddchannel}), the spreading functions of ESDD channels are discretized with the delay resolution $1/\mathbb W$ and the Doppler resolution  $1/\mathbb T$, corresponding to the signal's sampling rate and duration, respectively.
As a result, although the general LTV or DD channels do not have common properties, the corresponding ESDD channels do have common delay and Doppler resolutions, which are determined either by the signal or by the system.

Since the delay and Doppler have the same unit with the time and frequency, respectively, we consider MC modulation in the DD domain with the delay resolution of $1/\mathbb W$ and the Doppler resolution of $1/\mathbb T$, and design the corresponding pulse.

\subsection{DD Domain 2D Impulse Response}
The discretized spreading function in (\ref{esddchannel}) can be viewed as the ESDD channel's 2D impulse response, which \emph{looks like} being time-invariant.
Recall that the {gridded} DD domain has the physical units of time and frequency, so it is also a {gridded} TF domain.
As we will show below, the complex ICI of OFDM or OFDM/OQAM in LTV channels is exactly due to the
mismatch between the modulation and the channel.

Let us consider passing an MC signal $x(t)$ through the ESDD channel of (\ref{esddchannel}). The received waveform is given by
\begin{align}\label{yt}
  y(t)=\sum_{p=1}^P h_p x(t-\tau_p)e^{j2\pi \nu_p (t-\tau_p)},
\end{align}
where $x(t)=\sum_{m=0}^{M-1}x_m(t)$ as defined in (\ref{xt}) and (\ref{xmt}).
Assume that we set a large enough $\Delta T\ge T_g+\tau_{\textrm{max}}$ for completely isolating MC symbols from each other and preventing ISI. Then, from (\ref{yt}), the component of $y(t)$ corresponding to the $m$th MC symbol $x_m(t)$ is given by
\begin{align}\label{ymt}
  y_m(t) = & \sum_{p=1}^P h_p x_m(t-\tau_p)e^{j2\pi \nu_p (t-\tau_p)} \nonumber                                     \\
  =        & \sum_{p=1}^P h_p\sum_{n=-N/2}^{N/2-1} X[m,n]g(t-m\Delta T-\tau_p) \nonumber                            \\
           & \times e^{j2\pi n \Delta F (t-m\Delta T-\tau_p)}e^{j2\pi \nu_p (t-\tau_p)} \nonumber                   \\
  =        & \sum_{p=1}^P h_p e^{j2\pi \nu_p m\Delta T} \sum_{n=-N/2}^{N/2-1} X[m,n]g(t-m\Delta T-\tau_p) \nonumber \\
           & \times e^{j2\pi n \Delta F (t-m\Delta T-\tau_p)}e^{j2\pi \nu_p (t-m\Delta T-\tau_p)},
\end{align}
which is contaminated by ICI due to the Doppler-induced frequency dispersion. A well-known interpretation of this result is to consider $\nu_p$ as a \ac{cfo}, {as in conventional OFDM systems}. {Since} usually $\nu_p/\Delta F \not\in \mathbb Z$, this kind of \emph{fractional} CFO will cause severe ICI \cite{moose94}.

Eq. (\ref{ymt}) implies that for each $x_m(t)$, the ESDD channel is
\begin{equation}\label{ddm}
  h_m(\tau,\nu)=\sum_{p=1}^{P}h_p e^{j2\pi \nu_p m\Delta T} \delta (\tau-\tau_p)\delta (\nu-\nu_p),
\end{equation}
rather than $h(\tau,\nu)$ of (\ref{esddchannel}). In fact, as long as we have a symbol interval of $\Delta T >0$, (\ref{ddm}) holds regardless of whether the ISI exists or not. In the context of OFDM, the phase term
$e^{j2\pi \nu_p m\Delta T}$ in (\ref{ddm}) {has similar effects} as the phase difference between common phase errors of two adjacent OFDM symbols corrupted by the same CFO \cite{moose94}.
As a result, for a signal consisting of time-multiplexed symbols, different symbols experience different ESDD channels, {where these ESDD channels have the same number of paths, the delay and Doppler shifts, but different path gains.}
In other words, from the perspective of signal transmission, the 2D impulse response of an ESDD channel is \emph{still time-varying}, due to the phase term discussed above. Similar results can also be obtained straightforwardly for the DD channel in (\ref{sf}).

\subsection{Transmission Strategy for the ESDD Channel}
Bearing in mind that $\tau_p=l_p/\mathbb W$, $\nu_p=k_p/\mathbb T$, and $l_p, k_p \in \mathbb Z$ in (\ref{esddchannel}),  one can see from (\ref{ymt}) that if $\Delta F=1/\mathbb T$, we have $\nu_p/\Delta F =k_p\in \mathbb Z$, and then the ICI becomes aligned with the fine frequency resolution $1/\mathbb T$, which has a similar effect to that of
\emph{integer} CFO \cite{moose94}. The same thing happens to the ISI if $\Delta T=1/\mathbb W$.
In other words, if we can design a DDMC modulation as
\begin{align}\label{xtddmc}
  x(t)=\sum_{m=0}^{M-1}\sum_{n=-N/2}^{N/2-1} X[m,n]g\left(t-\frac{m}{\mathbb W}\right)e^{j2\pi n \frac{1}{\mathbb T} (t-\frac{m}{\mathbb W})},
\end{align}
the time and frequency resolutions of which are identical to those of the ESDD channel,
the ISI and ICI will be aligned to the ESDD channel grid. {As a result, each transmitted symbol is only interfered by a minimum number of its neighbor symbols, and the pattern of the whole interference become \emph{compact}.}
Further considering that the delay resolution $1/\mathbb W$ and the Doppler resolution $1/\mathbb T$ are common properties of ESDD channels, the DDMC modulation in (\ref{xtddmc}) seems to be an appropriate modulation scheme for the ESDD channel.

The aligned ISI and ICI can also be interpreted using the relation between the STFT of $x(t)$ and $y(t)$ in (\ref{stft_io}). Note that the STFT is equivalent to applying receive pulses (analysis window) to extract the signal components in {the} MC modulation. Similar to (\ref{rp}), upon substituting $\mathcal S(\tau,\nu)=h(\tau,\nu)$ and $t=m/\mathbb W$, $f=n/\mathbb T$  into the \ac{stft} in (\ref{stft_io}) to extract the signal component at  the $(m,n)$-th TF grid point, we have
\begin{align}\label{stft_io_discrete1}
  Y[m,n] = & Y^{(g)}\left(\frac{m}{\mathbb W},\frac{n}{\mathbb T}\right)  \nonumber                                                                                \\
  =        & \sum_{p=1}^P h\left(\frac{l_p}{\mathbb W},\frac{k_p}{\mathbb T}\right) X^{(g)}\left(\frac{m-l_p}{\mathbb W}, \frac{n-k_p}{\mathbb T}\right) \nonumber \\
           & \times e^{-j2\pi \frac{l_p(n-k_p)}{\mathbb W\mathbb T}}.
\end{align}
Then, we have the following proposition.
\begin{prop}\label{l0}
  If the transmit pulse $g(t)$ is an orthogonal pulse with respect to {the delay resolution $1/\mathbb W$ and Doppler resolution $1/\mathbb T$ of the ESDD channel} as defined in (\ref{g_ortho}), the received signal component at the $(m,n)$-th TF grid point obeys
  \begin{align}\label{stft_io_discrete2}
    Y[m,n]
    = & \sum_{p=1}^P h\left(\frac{l_p}{\mathbb W},\frac{k_p}{\mathbb T}\right) X[m-l_p,n-k_p]
    \nonumber                                                                                 \\
      & \times  e^{-j2\pi \frac{l_p(n-k_p)}{\mathbb W\mathbb T}}.
  \end{align}
\end{prop}
\begin{IEEEproof}
  By substituting the orthogonality property of $g(t)$ into (\ref{stft_io_discrete1}), (\ref{stft_io_discrete2}) can be obtained straightforwardly.
\end{IEEEproof}

Proposition \ref{l0} reveals the basic form of the IO relation for DDMC over the ESDD channel.
From (\ref{stft_io_discrete2}), one can see that \emph{except for some phase terms}, the extracted signal component at each TF grid point in the DDMC can be described as a 2D convolution between the transmitted digital symbols and the ESDD channel. Upon considering the impulse-function-based transmission strategy of SC modulation and the resultant 1D convolutional IO relation over the LTI channel, it becomes clear that the DDMC modulation is an \emph{impulse-function-based transmission strategy for the ESDD channel}.

It is apparent that an appropriately designed $g(t)$ is necessary to realize such an impulse-function-based transmission. However, the impulse in the DD domain or DDLP has a TFA less than $1/(4\pi)$, {which violates} the Heisenberg uncertainty principle. Therefore, DDLP does not exist\cite{hadaniyt}. Meanwhile, because the JTFR is now $\Delta R_{\textrm{DD}}=1/(\mathbb W \mathbb T)\ll 1$, according to the WH frame theory, (bi)orthogonal WH sets do not exist either. Hence, it seems impossible to perform DD domain modulation, due to the lack of pulses.
In the following sections, we will discuss progress in the design of DD domain modulation.

\subsection{OTFS Modulation}
Modulating information-bearing symbols in the DD domain was first considered in form of the OTFS modulation \cite{otfs_wcnc_2017,hadani_otfs_2018}.
To avoid the aforementioned pulse design challenges, the OTFS modulation transforms the signals from the DD domain to the TF domain by the ISFFT precoder. Then it modulates the transformed {signals} using the symbol-wise CP-free OFDM, and prepends a frame based CP for the whole OTFS frame.

Let the frequency resolution (subcarrier spacing) of the CP-free OFDM be $F_0$, the resulting symbol period becomes $T_0=1/F_0$. Then, the TF domain grid of OTFS obeys $\{\hat nT_0,\hat m\frac{1}{T_0}\}$ for $\hat n=0,\ldots,N-1$ and $\hat m=0,\ldots M-1$, while the corresponding DD domain grid is defined as $\left\{m\frac{T_0}{M}, n\frac{1}{NT_0}\right\}$ for $m=0,\ldots,M-1$ and $n=0,\ldots N-1$.
Then, the waveform of an OTFS frame without the frame-wise CP can be written as\cite{otfs_wcnc_2017}
\begin{align}\label{xtotfs}
  \hat x(t)=\sum_{\hat n=0}^{N-1}\sum_{\hat m=-\frac{M}{2}}^{\frac{M}{2}-1} \mathcal X[\hat n,[\hat m]_M] g(t-\hat nT_0)e^{j2\pi \hat m F_0 (t-\hat n T_0)},
\end{align}
where $\mathcal X[\hat n,\hat m]$ is obtained by the ISFFT as
\begin{equation}\label{isfft}
  \mathcal X[\hat n,\hat m] = \frac{1}{\sqrt{MN}} \sum_{m=0}^{M-1} \sum_{n=0}^{N-1} X[m,n]e^{j2\pi(\frac{\hat n n}{N}-\frac{\hat mm}{M})}.
\end{equation}
Note that in OTFS, $M$ and $N$ are also {the number of subcarriers and symbols of the underlying OFDM modulation}, respectively.
Observe from (\ref{xtotfs}), the OTFS waveform is designed exactly with $\Delta R=\Delta T \Delta F=1$, because $\Delta T= T=T_0$.

  {The ideal pulse of OTFS in (\ref{xtotfs}) is said to satisfy the \emph{biorthogonal robustness property}\cite{otfs_wcnc_2017}.
    Roughly speaking, this means that the pulses $g(t-\hat nT_0)e^{j2\pi \hat m F_0 (t-\hat n T_0)}$ for different $m$ or $n$ in (\ref{xtotfs}) are (bi)orthogonal with each other, even after experiencing the time and frequency dispersion induced by the channel. }
Given this ideal pulse, OTFS expects to achieve a 2D convolution between the transmit symbols $X[m,n]$ and the channel $h(\tau, \nu)$ at the channel output\cite{hadani_otfs_2018}.
Unfortunately, the assumed ideal pulse \emph{cannot} be realized in practice\cite{hadani_otfs_2018}. The underlying reason may be that the pulses satisfying the biorthogonal robustness property essentially correspond to the eigenfunction-based transmission that achieves a scalar IO relation over the LTV channel, which requires $\Delta R>1$, as mentioned in Section \ref{tfmc_for_ltv}.
Since the OTFS waveform is designed for $\Delta R=1$, the corresponding TLOP cannot be biorthogonal robust.

Due to the absence of ideal pulse, the rectangular pulse $\Pi_{T_0}(t)$ has been widely adopted in current OTFS studies\cite{viterbo_twc_2018}, which is the TLOP for the CP-free OFDM with $\Delta R=1$. Recall that for the CP-free OFDM, some vacant edge subcarriers are necessary to suppress the OOBE, as mentioned in Section \ref{unload_subcarrier}. Since OTFS relies on CP-free OFDM, OOBE is also an inevitable practical issue for OTFS \cite{shen2022error}.
Furthermore, the severe bandwidth leakage invalidates the intended nominal bandwidth $\frac{M}{T_0}$ and the associated delay resolution $\frac{T_0}{M}$, which the OTFS signal was designed to align with.
However, letting $\mathcal X[\hat n,\hat m]=0$ for a part of $\hat m, 0\le \hat m \le M-1$ may break the inherent connection between $X[m, n]$ and $\mathcal X[\hat n,\hat m]$ governed by the ISFFT precoder in (\ref{isfft}), as $X[m, n]$ is drawn from a QAM constellation.
Also, the absence of CP and CS in CP-free OFDM makes the windowing-based OOBE mitigation methods infeasible, while letting $g(t)$ in (\ref{xtotfs}) be a spectrally compact pulse leads to severe performance degradation \cite{viterbo_tvt_19} due to loss of orthogonality.
Furthermore, it is noteworthy that from (\ref{ddm}),
the expected 2D convolution may be \emph{unachievable}, considering the phase terms induced by the time-varying path gains.

\subsection{DDMC/ODDM Modulation}
Motivated by OTFS's concepts of modulating information-bearing symbols in the DD domain and DD grid, we present a general DDMC signal design in this section. Considering the DD domain grid, let us substitute $\frac{1}{\mathbb W}=\frac{T_0}{M}$ and $\frac{1}{\mathbb T}=\frac{1}{NT_0}$ into (\ref{xtddmc}), and replace the prototype pulse $g(t)$ with $u(t)$, for the sake of comparison. Then, the DDMC modulation waveform becomes
\begin{align}\label{xtoddm}
  x(t) & =\sum_{m=0}^{M-1}\sum_{n=-N/2}^{N/2-1} X[m,n]u\left(t-m\frac{T_0}{M}\right)e^{j2\pi n \frac{1}{NT_0} (t-m\frac{T_0}{M})}, \nonumber \\
       & =\sum_{m=0}^{M-1}\sum_{n=-N/2}^{N/2-1} X[m,n]u_{m,n}(t),
\end{align}
where the time and frequency resolutions are $\Delta T=\frac{T_0}{M}=\frac{1}{MF_0}$ and $\Delta F=\frac{1}{NT_0}=\frac{F_0}{N}$, respectively.

Now the crucial question is whether the DDOP denoted by $u(t)$ in (\ref{xtoddm}), which is orthogonal with respect to both the time (delay) resolution $\Delta T=\frac{T_0}{M}$ and the frequency (Doppler) resolution $\Delta F=\frac{1}{NT_0}$, exists or not.
If such a DDOP does exist, we can realize this hypothetical ODDM modulation.
It turns out that although there does not exist $g(t)$ that satisfies the orthogonality condition (\ref{g_ortho}) for $\Delta T=\frac{T_0}{M}$ and $\Delta F=\frac{1}{NT_0}$ across all $m$ and $n$, if we limit their range, we can have:
\begin{enumerate}
  \item orthogonality condition within the range holds,
  \item the \ac{io} relation in Proposition 1 holds within the same range.
\end{enumerate}
As explained in \cite{oddmicc22,oddm}, the DDOP that satisfies this new orthogonality condition with limited range is
a time-limited or windowed \emph{pulse-train} given by
\begin{equation}\label{ut}
  u(t)=\sum_{\dot n=0}^{N-1}a(t-\dot nT_0),
\end{equation}
where the subpulse $a(t)$ is a square-root Nyquist pulse \emph{parameterized} by its zero-ISI namely Nyquist interval $\frac{T_0}{M}$ and its duration $T_a=2Q\frac{T_0}{M}$, {where $Q$ can be a positive integer}.

Fig. \ref{shape_ut} illustrates the unique structure of the DDOP $u(t)$. Note that the orthogonality property of a prototype pulse in MC modulation, such as $u(t)$, can be understood in terms of its ambiguity function. In particular, when $2Q\ll M$ and therefore $T_a\ll T_0$, it has been proved that $u(t)$ satisfies the orthogonality property of
\begin{equation}
  \langle u(t), u_{m,n}(t)\rangle = \mathcal A_{u,u}\left(m\frac{T_0}{M}, n\frac{1}{NT_0}\right)= \delta(m)\delta(n),\label{orth}
\end{equation}
for $|m|\le M-1$ and $|n| \le N-1$, where $\mathcal A_{u,u}(\cdot)$ is the ambiguity function of $u(t)$ defined as
\begin{align}
  \mathcal A_{u,u}(\tau,\nu) & =   \langle u(t), u(t-\tau)e^{j2\pi \nu (t-\tau)} \rangle,           \nonumber \\
                             & =  \int_{-\infty}^{\infty} u(t) u^{*} (t-\tau)e^{-j2\pi \nu (t-\tau)} dt.
\end{align}
In other words, $u(t)$ is orthogonal with respect to the {delay and Doppler resolutions of the ESDD channel within the range of interest, $|m|\leq M-1$ and $|n|\leq N-1$}. Intuitively, the time resolution of this DDOP $u(t)$  is captured by the sharpness of each component $a(t)$, while the Doppler resolution is captured by the periodic structure of the overall pulse train. The orthogonality property of $u(t)$ can be shown by analyzing its time and frequency shifted version within their respective ranges.

\begin{figure}
  \centering
  \includegraphics[width=8cm]{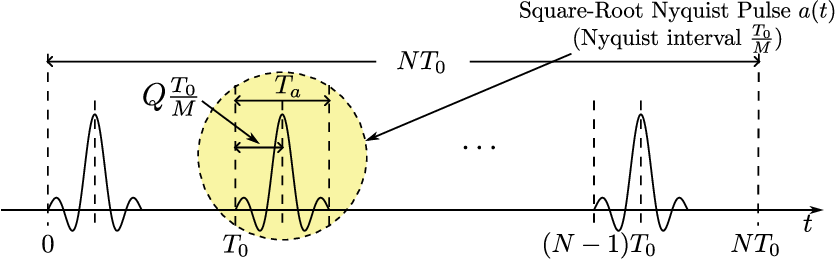}
  \caption{Delay Doppler domain orthogonal pulse (DDOP) $u(t)$.}
  \label{shape_ut}
\end{figure}

After prepending an appropriate frame based CP to $x(t)$ in (\ref{xtoddm}), the ODDM frame is sent through the channel.
Let us assume that the maximum delay and Doppler of the ESDD channel in (\ref{esddchannel}) are  $(L-1)\frac{T_0}{M}$ and $K\frac{1}{NT_0}$, respectively\footnote{Understanding the orthogonality characteristics of DDOP in physical channels with fractional/off-grid delay and Doppler is crucial for determining how well DDOP couples with such channels. For further insights, please refer to \cite{2024_TCOM_JunTong_ODDMoverPhysicalChannels}.}.
The $P$ paths can be arranged in a $(2K+1)\times L$ DD domain channel matrix $\mathbf \Theta$, where each row and each column of $\mathbf \Theta$ correspond to a Doppler and delay index, respectively. For example, let $\hat k=k-K-1$, a nonzero element of $\mathbf \Theta$, denoted by $\theta(\hat k+K+1,l)$, be equal to the gain $h_p$ of the $p$th path, whose delay and Doppler are $l\frac{T_0}{M}$ and $\hat k \frac{1}{NT_0}$, respectively. Note that the total number of nonzero elements in $\mathbf \Theta$ is $P$.

At the receiver, a matched filtering (or correlator) operation based on $u(t-m\frac{T_0}{M})e^{-j2\pi n \frac{1}{NT_0}(t-m\frac{T_0}{M})}$ is performed to obtain $Y[m,n]$, the signal component at the $(m,n)$-th TF grid point,  namely at the $n$-th subcarrier of the $m$-th ODDM symbol. Let us arrange $Y[m,n]$ and $X[m,n]$ in two vectors as
\begin{align*}
  \mathbf y  =  [\mathbf y_0^T,\mathbf y_1^T, \cdots, \mathbf y_{M-1}^T ]^T, \\
  \mathbf x  =  [\mathbf x_0^T,\mathbf x_1^T, \cdots, \mathbf x_{M-1}^T ]^T,
\end{align*}
where
\begin{align*}
  \mathbf y_m & =[Y[m,0],Y[m,1],\cdots,Y[m,N-1]]^T, \\
  \mathbf x_m & =[X[m,0],X[m,1],\cdots,X[m,N-1]]^T,
\end{align*}
for $0\le m \le M-1$.
Then, the noise-free IO relation of ODDM over the ESDD channel can be written as
\begin{equation} \label{ioddcompact}
  \mathbf y=\mathbf H \mathbf x,
\end{equation}
where the DD domain channel matrix $\mathbf H$ is given by
\small
\begin{equation}\label{H}
  \mathbf H \triangleq
  \begin{bmatrix}
    \mathbf H_0^0         &        &        &                       & \mathbf H_{L-1}^0 \mathbf D & \cdots         & \cdots & \mathbf H_{1}^0  \mathbf D      \\
    \vdots                & \ddots &        &                       &                             & \ddots         & \ddots & \vdots                          \\
    \vdots                & \ddots & \ddots &                       &                             &                & \ddots & \vdots                          \\
    \mathbf H_{L-2}^{L-2} & \ddots & \ddots & \mathbf H_{0}^{L-2}   &                             & \mbox{\Huge 0} &        & \mathbf H_{L-1}^{L-2} \mathbf D \\
    \mathbf H_{L-1}^{L-1} & \ddots & \ddots & \ddots                & \mathbf H_{0}^{L-1}         &                                                           \\
                          & \ddots & \ddots & \ddots                & \ddots                      & \ddots                                                    \\
                          &        & \ddots & \ddots                & \ddots                      & \ddots         & \ddots                                   \\
    \mbox{\Huge 0}        &        &        & \mathbf H_{L-1}^{M-1} & \ddots                      & \ddots         & \ddots & \mathbf H_{0}^{M-1}
  \end{bmatrix}
\end{equation}
\normalsize
with
\begin{eqnarray}
  \mathbf H_l^m & = & \sum_{\hat k=-K}^{K} \theta(\hat k+K+1,l) e^{j2\pi \frac{\hat k(m-l)}{MN} } \mathbf C^{\hat k}, \\
  \mathbf D & = & \textrm{diag}\left\{1, e^{-j\frac{2\pi}{N}},\ldots, e^{-j\frac{2\pi(N-1)}{N}}\right\},
\end{eqnarray}
and the $N \times N$ cyclic permutation matrix is formulated as
\begin{equation}
  \mathbf C=
  \begin{bmatrix}
    0      & \ldots & 0      & 1      \\
    1      & \ddots & 0      & 0      \\
    \vdots & \ddots & \ddots & \vdots \\
    0      & \ldots & 1      & 0
  \end{bmatrix}.
\end{equation}
As an  $MN \times MN$ block-circulant-like matrix, the DD domain channel matrix $\mathbf H$ in (\ref{ioddcompact}) represents the linear combination between {$X[m,n]$} and the ESDD channel $h(\tau, \nu)$.
Upon using the DDOP $u(t)$ of Fig. \ref{shape_ut} as the transmit and receive prototype pulses, ODDM becomes capable of outperforming the OTFS in terms of both its OOBE and \ac{ber}\cite{oddmicc22,oddm}.
It is noteworthy that, as observed in (\ref{xtoddm}), the information-bearing symbols are first modulated onto carriers and subsequently aggregated. Then, each ODDM symbol is obtained by applying truncation or transmit pulse shaping to the aggregated symbols using the DDOP
$u(t)$, which is a pulse train. Therefore, ODDM can be interpreted as a \ac{psofdm} or more precisely a \ac{ptsofdm}.
The orthogonality of the DDOP with respect to the DD resolution in ODDM ensures that signals transmitted for different $X[m,n]$ are mutually orthogonal to each other. Furthermore, it ensures
\emph{a sparse interference pattern}
after transmission over the ESDD channels,
because each path of the channel contributes exactly one of the transmitted symbol at the receiver of ODDM, which can be verified with $P=1$ and the resultant permuted diagonal matrix $\mathbf H$ in (\ref{H}).

The existence of the DDOP with orthogonality in (\ref{orth}) comes as a bit of surprise, because the conventional understanding of (bi)orthogonal pulses indicates that orthogonal WH sets do not exist along with such fine JTFRs as $\Delta R_{\textrm{DD}}=  T_0/M\times 1/(NT_0)=1/(MN)\ll 1$, in line with classic WH frame theory. However, the WH frame theory is for full WH sets, and it is not necessarily applicable to WH subsets, which are used in practical modulation schemes having limited time-frequency resources. The existence of $u(t)$ thus motivates us to reconsider the fundamental (bi)orthogonal pulse design principles based on WH frame theory, which will be discussed in the next section.

To summarize, the key motivations for ODDM are that: (i) information is modulated directly in the \ac{dd} domain; (ii) similar to (\ref{g_ortho}) and (\ref{g_biortho}), the \ac{ddop} is (bi)orthogonal with respect to the delay resolution $\frac{T_0}{M}$ and the Doppler resolution $\frac{1}{NT_0}$ within finite TF shifts. Although the corresponding  (bi)orthogonality property holds only within a finite TF region with a range  of $(T_0, \frac{1}{T_0})$, the \ac{io} relationship becomes much simpler, thereby allowing significant complexity saving in channel equalization.

\def\arraystretch{1.2}
\begin{table}[t]
  \caption{Main system parameters of ODDM and \\ Underlying OFDM in OTFS}
  \begin{center}
    \begin{tabular}{|c|c|c|}
      \hline                                 & ODDM             & Underlying OFDM in OTFS   \\
      \hline frequency resolution $\Delta F$ & $\frac{1}{NT_0}$ & $\frac{1}{T_0}$           \\
      \hline time resolution $\Delta T$      & $\frac{T_0}{M}$  & $T_0$                     \\
      \hline Number of subcarriers           & $N$              & $M$                       \\
      \hline Number of MC symbols            & $M$              & $N$                       \\
      \hline Prototype pulse                 & $u(t)$           & $\Pi_{T_0}\left(t\right)$ \\
      \hline
    \end{tabular}\label{tab:uOFDMvsODDM}
  \end{center}
\end{table}

\subsection{ODDM versus OTFS}

As final remark in this section, we highlight the differences between \ac{oddm}\cite{oddm} and \ac{otfs}\cite{otfs_wcnc_2017}. One can see from (\ref{xtoddm}) and (\ref{xtotfs}), they are two fundamentally different MC waveforms, where the core differences lie in their TF resolution, the resulting orthogonality and basis functions.

In particular, ODDM uses an MC modulator with fine TF or DD resolution, given by $\left(\frac{T_0}{M}, \frac{1}{N T_0}\right)$, where a set of mutually orthogonal transmit pulses is obtained by applying TF shifts to the prototype pulse $u(t)$ according to the DD resolution. In contrast, OTFS employs an ISFFT precoder followed by a conventional TFMC or OFDM modulator with a coarser TF resolution $\left(T_0, \frac{1}{T_0}\right)$. As a result,  in OTFS, the \emph{effective} transmit pulses corresponding to the original DD domain QAM symbols, though still mutually orthogonal, are \emph{not} orthogonal with respect to the DD resolution $\left( \frac{T_0}{M}, \frac{1}{N T_0} \right)$ \cite{2024_TCOM_JunTong_ODDMoverPhysicalChannels}. In other words, one effective transmit pulse in OTFS cannot be obtained by applying TF shifts to another pulse with respect to the DD resolution.

The main system parameters of the underlying OFDM in OTFS and ODDM are summarized in Table \ref{tab:uOFDMvsODDM}. A comprehensive comparison of ODDM and OTFS
\footnote{We note that the modulation and multiplexing schemes recently designed for LTV channels, such as \cite{LTVDesign_1,LTVDesign_3,LTVDesign_4}, share notable similarities and differences with ODDM. Specially, the concept of coupling information-bearing symbols with the DD domain representation of the channel was first introduced with OTFS and later explored with ODDM. Moreover, ODDM first proposed an analog orthogonal pulse specifically catered for DDMC in the form of DDOP, while other waveform designs are provided in the form of digital sequences. Given these similarities and differences, it is important to provide a thorough comparison between these designs and ODDM. This will be addressed in detail in our future works.}
\footnote{It is worth noting that ODDM is defined by a unique set of orthogonal basis functions under the evolved WH framework\cite{primer}. On the other hand, the concept of Zak-OTFS was initially proposed using the discrete Zak transform to generate a discrete-time sequence\cite{hadaniyt}, and OTFS 2.0 further incorporates pulse-shaping into this scheme\cite{otfsbits}, which, to the best of our knowledge, does not naturally fall within the WH framework. As these represent distinct lines of development, this paper focuses its comparison on the original ISFFT and OFDM-based OTFS to ensure a fair evaluation within a common WH-theoretic framework. A comprehensive comparison with other members of the OTFS family or other DD modulation schemes from a basis function perspective is an important topic deserving of future investigation.}, covering their basis functions (or effective transmit pulses), \ac{oobe}, and orthogonality, can be found in \cite{2024_TCOM_JunTong_ODDMoverPhysicalChannels}.

\section{Pulse Design Subject to the Signal's TF Constraints}
As can be observed from (\ref{orth}), the (bi)orthogonality of \ac{oddm} is constrained within
$M$ symbols, each having $N$ subcarriers. Therefore it only applies
to a local region in the TF domain. Since MC modulation has a limited
number of symbols and subcarriers, the orthogonality within
this local TF region defined by the bandwidth and duration of the signal is sufficient. As a result, we can reformulate the pulse design problem for MC modulation by taking the {TF constraints of the practical signal} into account.

Without loss of generality, let us consider the TF region in Fig. \ref{tf_lattice}, where the bandwidth and duration are $B_x$ and $T_x$, respectively.
It is widely understood that for a signal contained in this region, its DoF is bounded by $B_xT_x$, which can be achieved {by} using the \ac{psw} functions \cite{landau62}. In other words, we can transmit up to $\ceil{B_xT_x}$ digital symbols, by carrying them using the PSW functions. However, the PSW functions
neither have a complex sinusoidal based structure required by MC {modulations, nor can they} benefit the channel equalization. Upon further considering the dispersive channel effects and the DoF of the received signal, it might be necessary to relax the sampling rate $\mathbb W$ and samples duration $\mathbb T$ to $\mathbb W \le B_x$ and/or $\mathbb T \le T_x$ to design an MC modulation associated with other (bi)orthogonal functions and transmit up to $\ceil{\mathbb W\mathbb T}\le \ceil{B_xT_x}$ digital symbols, at the cost of a modest erosion of bandwidth efficiency.

\subsection{Global and Local/Sufficient (Bi)orthogonality}
Analogous to (\ref{g_ortho}) and (\ref{g_biortho}), the (bi)orthogonal pulse design problem taking the {TF constraints of the signal, namely the limited number of symbols and subcarriers,} into account is
to find specific WH \emph{subsets} $ \left(g,\Delta T, \Delta F, M, N\right)$ and $ \left(\gamma ,\Delta T, \Delta F, M, N\right)$ that satisfy the orthogonal condition of
\begin{equation}\label{l_ortho}
  \langle g_{m,n}, g_{\dot m,\dot n}\rangle =\delta(m-\dot m)\delta(n-\dot n),\,\, m,\dot m\in \mathbb Z_M, n,\dot n\in \mathbb Z_N,
\end{equation}
or the biorthogonal condition of
\begin{equation}\label{l_biortho}
  \langle g_{m,n}, \gamma_{\dot m,\dot n}\rangle =\delta(m-\dot m)\delta(n-\dot n),\,\, m,\dot m\in \mathbb Z_M, n,\dot n\in \mathbb Z_N,
\end{equation}
where
\begin{align}
  \mathbb Z_M =\{0, \cdots, M-1\}, \,\, \mathbb Z_N =\{0, \cdots, N-1\}.
\end{align}
Here, the index of subcarriers $\{-N/2,\cdots\, 0,\cdots, N/2-1\}$ is changed to $\{0,\cdots, N-1\}$ for simplifying the notation, which corresponds to a half-bandwidth shift of the carrier frequency $f_c$ \cite{oddmicc22}. This will not affect the analysis of (bi)orthogonality.

Since (\ref{l_ortho}) and (\ref{l_biortho}) only consider a local region in the TF domain, we term them as the {\emph{local/sufficient orthogonal
      condition} and \emph{local/sufficient biorthogonal condition}}, respectively.
Furthermore, because of
\begin{align}
  \langle g_{m,n}, g_{\dot m,\dot n}\rangle
  = \mathcal A_{g,g} (\bar m \Delta T, \bar n \Delta F)e^{j2\pi n\bar m \Delta F \Delta T},
\end{align}
where $\bar m= \dot m-m$ and $\bar n= \dot n-n$, the local/sufficient orthogonal condition in (\ref{l_ortho}) is equivalent to
\begin{align}\label{agg}
  \mathcal A_{g,g}(\bar m\Delta T, \bar n\Delta F)=\delta(\bar m)\delta(\bar n),
\end{align}
for $|\bar m| \le M-1, |\bar n| \le N-1$. Similar results can be obtained for the local/sufficient biorthogonal condition in (\ref{l_biortho}). 

In the context of TFA, WH sets are used for analyzing finite-energy signals lying in the space of $L^2(\mathbb R)$. For accurate analysis, the WH sets have to be WH frames, which are complete or overcomplete WH sets with a certain guaranteed numerical stability of reconstruction and this requires $\Delta R\le 1$\cite{gaborana,ftfa}.
When a WH set $\left(g,\Delta T, \Delta F\right)$ is a WH frame, we can denote it as $\left\{g,\Delta T, \Delta F\right\}$ by replacing round brackets with curly brackets.
Let $\Delta T_{\dagger}= 1/\Delta F$ and $\Delta F_{\dagger}= 1/\Delta T$.
From the duality and biorthogonality theory for WH frames\cite{wexler1990,janssen1995,daubechies95}, we know that {$\left(g,\Delta T, \Delta F\right)$} and {$\left(\gamma,\Delta T, \Delta F\right)$} are biorthogonal if and only if the associated WH sets $(g,\Delta T_{\dagger}, \Delta F_{\dagger})$ and $(\gamma,\Delta T_{\dagger}, \Delta F_{\dagger})$ are dual frames,
while {$\left(g,\Delta T, \Delta F\right)$} is orthogonal if and only if the associated WH set $(g,\Delta T_{\dagger}, \Delta F_{\dagger})$ is a tight frame\footnote{A WH frame is called tight if its lower and upper frame bounds are the same.}.
To obtain the dual frames $\left\{g,\Delta T_{\dagger}, \Delta F_{\dagger}\right\}$ and  $\left\{\gamma,\Delta T_{\dagger}, \Delta F_{\dagger}\right\}$ or the tight frame
$\left\{g,\Delta T_{\dagger}, \Delta F_{\dagger}\right\}$, the corresponding JTFR  has to satisfy $\Delta R_{\dagger}=\Delta T_{\dagger}\Delta F_{\dagger} \le 1$ and consequently $\Delta R=\Delta T \Delta F \ge 1$. Therefore, (bi)orthogonal WH sets do not exist for $\Delta R <1$.

The WH frame theory based results regarding (bi)orthogonal WH sets are rigorous. However, since a WH set is originally a TFA tool conceived for functions in $L^2(\mathbb R)$, it considers the whole TF domain where $m,n\in \mathbb Z$, and corresponds to the signal without the limit of bandwidth and duration.
To make this possible, given $\Delta T$ and $\Delta F$, $g(t)$ must be \emph{independent} of the number of symbols $M$ and the number of subcarriers $N$, to be shifted freely over the whole TF domain. As a result, $g(t)$ is designed based only on $\Delta F$ and $\Delta T$ to achieve the \emph{global} (bi)orthogonality in (\ref{g_ortho}) and (\ref{g_biortho}), and therefore bounded by the JTFR limit of $\Delta R=1$.

  {On the other hand, for MC {modulations}, in contrast to achieving the \emph{global} (bi)orthogonality in (\ref{g_ortho}) and (\ref{g_biortho}), we only have to consider the \emph{local/sufficient} (bi)orthogonality in (\ref{l_ortho}) and (\ref{l_biortho}) to satisfy the perfect reconstruction condition for $M$ MC symbols with $N$ subcarriers, corresponding to a WH \emph{subset} under the {TF constraints of the signal}.
    Apparently, $g(t)$ that achieves the global (bi)orthogonality can form a (bi)orthogonal WH subset. However, since we really only require a WH subset to satisfy the local/sufficient (bi)orthogonality, it is not necessarily bound by the WH
    frame theory for the WH set. In fact, the pulses parameterized
    by not only $\Delta T$ and $\Delta F$ but also by $M$ and $N$ can also achieve the local/sufficient orthogonality.}

\subsection{Orthogonality w.r.t. $\Delta F$}

Let us consider a fixed $m$ and variable $n$ in $g_{m,n}$, and investigate the orthogonality with respect to the frequency resolution $\Delta F$ first. We want to find $g(t)$ that can achieve the orthogonality among $g(t-m\Delta T)e^{j2\pi n \Delta F (t-m\Delta T)}$ with a given $m$ but variable $n$, where $0\le  t \le T_g$ and $T_g= T=1/\Delta F$. Without loss of generality, let $m=0$, we can obtain the following results:
\renewcommand{\labelenumi}{F\arabic{enumi})}
\begin{enumerate}
  \item Unbounded $n$ ($n\in \mathbb Z$):
        $g(t)$ is the rectangular pulse $\Pi_T(t)$, which is independent of $N$.

  \item Bounded $n$ ($|n|\le N-1$): We have the following proposition:
\end{enumerate}
\begin{prop}\label{l1}
  When $g(t)$ is a {unit energy} periodic function with a period of $\frac{T}{N}$ for $0\le  t \le T_g$ and $T_g=T$, it satisfies the orthogonal property of
  \begin{equation}
    \mathcal A_{g,g}\left(0, n\Delta F\right)= \langle g(t), g(t)e^{j2\pi n\Delta Ft} \rangle=\delta(n).
  \end{equation}
  for $|n|\le N-1$.
\end{prop}
\begin{IEEEproof}
  See Appendix A.
\end{IEEEproof}
Proposition \ref{l1} indicates that once there is a constraint imposed on the number of subcarriers, surprisingly there are an \emph{infinite} number of pulses satisfying the orthogonality with respect to $\Delta F$ within the symbol period of $T=1/\Delta F$. In other words, as long as $g(t)$ is a periodic function satisfying the above conditions, \emph{regardless of its bandwidth $B_g$}, it can achieve the orthogonality among $N$ subcarriers.
Considering the case of F1) where $B_g$ is proportional to $\Delta F$ and the {total bandwidth of the signal} is about $N\Delta F$, Proposition \ref{l1} actually decouples the relation between $B_g$ and $\Delta F$. An example of such a function $g(t)$ for $N=4$ is shown in Fig. \ref{gtF2}.

\begin{figure}[t]
  \centering
  \includegraphics[width=7cm]{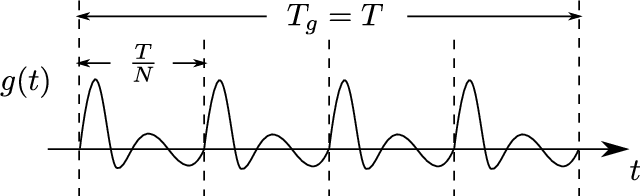}
  \caption{$g(t)$ orthogonal w.r.t $\Delta F=\frac{1}{T}$ for $|n|\le N-1$ and fixed $m$}
  \label{gtF2}
\end{figure}

\subsection{Orthogonality w.r.t. $\Delta T$}
Similarly, we can consider a fixed $n$ in $g_{m,n}$, and investigate the orthogonality with respect to the time resolution $\Delta T$.
Our target now becomes that of  finding a specific $g(t)$ that can achieve the orthogonality among $g(t-m\Delta T)e^{j2\pi n \Delta F (t-m\Delta T)}$ with a fixed $n$ but variable $m$. When $n\ne 0$, we have the following straightforward answer with  \emph{temporally isolated} pulses/subpulses:
\renewcommand{\labelenumi}{T\arabic{enumi})}
\begin{enumerate}
  \item Unbounded $m$ ($m\in \mathbb Z$) : Any $g(t)$ with duration $T_g\le \Delta T$, which is independent of $M$.
  \item Bounded $m$ ($|m|\le M-1$) : $g(t)$ is a pulse-train being made up of $\dot N >1 $ subpulses $b_{\dot n}(t), 0\le \dot n \le \dot N-1$, where these subpulses are temporally spaced by $M\Delta T$ and each subpulse has a duration $T_{b_{\dot n}} \le \Delta T$.
\end{enumerate}
Meanwhile, when $n=0$, we have another solution associated with \emph{temporally overlapped} pulse/subpulses:
\begin{enumerate}
  \setcounter{enumi}{2}
  \item Unbounded $m$ ($m\in \mathbb Z$) : square-root Nyquist pulse $a_{\Delta T}(t)$ {with $\Delta T$ being the zero-ISI or Nyquist interval}, which is also independent of $M$.
  \item Bounded $m$ ($|m|\le M-1$) : $g(t)$ is a pulse-train being made up of $\dot N >1 $ {square-root Nyquist subpulses $a_{\Delta T}(t)$, where these subpulses} are temporally spaced by $M\Delta T$.
\end{enumerate}

\subsection{Local/Sufficient Orthogonality w.r.t. $\Delta F$ and $\Delta T$}
From T1)-T4), we know that given the time resolution $\Delta T$, the key to achieving the orthogonality among $M$ symbols is to either limit the pulse duration to be {no} greater than $\Delta T$ or to employ square-root Nyquist pulses whose Nyquist interval is $\Delta T$. Also, for the case of the pulse-train in T2) and T4) corresponding to the local/sufficient orthogonality, the subpulses have to be temporally spaced by $M\Delta T$. At the same time, we know from F2) that given the frequency resolution $\Delta F$, the key to achieving the local/sufficient orthogonality among $N$ subcarriers is to form a periodic function with period $\frac{1}{N\Delta F}$. Therefore, to achieve the local/sufficient orthogonality with respect to $\Delta F$ and $\Delta T$ concurrently, we can consider a combination of the conditions in T2), T4) and F2). Furthermore, because the subpulses in T2) have much shorter duration and therefore much wider bandwidth than those in T4), the combination of T4) and F2) is preferred.

Given the time resolution $\Delta T$ and assuming that the duration of $a_{\Delta T}(t)$ obeys $T_{a_{\Delta T}} < M\Delta T$, it is interesting to observe that by letting the number of subpulses to be $\dot N=N$, the pulse-train in T4) is a periodic function that satisfies F2), if we let $T_g=T=MN\Delta T$ and $\Delta F =1/T=1/(MN\Delta T)$ in F2).
Further, upon substituting $\frac{T_0}{M}$ into $\Delta T$, the pulse-train becomes the DDOP $u(t)$ of (\ref{orth}) and achieves the orthogonality with respect to $\Delta T=\frac{T_0}{M}$ and $\Delta F =1/(N T_0)$, where $\Delta R =1/(MN) \ll 1$.

As a result, \emph{by combining the pulse-train structure required by the orthogonality with respect to the Doppler resolution, {and} the square-root Nyquist pulse required by the orthogonality with respect to the delay resolution}, we can bypass the JTFR limit of $\Delta R\ge 1$ for global (bi)orthogonality to achieve local/sufficient orthogonality with $\Delta R \ll 1$.
Compared to the traditional WH set based principles, the WH subset based principles and the resultant pulse-train structure may pave the way for conceiving new pulse designs for MC modulations.

\subsection{General DDOP and Local/Sufficient Biorthogonality}
The above result of sufficient orthogonality is based on the appropriate duration of $a_{\Delta T} (t)$ namely $T_{a_{\Delta T}} < M\Delta T$.
Recall that the orthogonality of the DDOP in (\ref{orth}) is also subject to a similar duration constraint of $T_a\ll T_0$ (equivalently $2Q\ll M$).
In practice, it is desirable to relax this constraint for the sake of flexible design. In the following, we will show that this duration constraint can be relaxed by introducing the cyclic extension, which leads to a general DDOP design.

When $T_a>T_0$, $u(t)$ is no longer a period function with a period of $T_0$ during $[0,NT_0]$, which is required to satisfy the orthogonality with respect to  $\Delta F=\frac{1}{NT_0}$.
This observation inspires us to use a cyclically extended version of $u(t)$ namely $u_{ce}(t)$, as the transmit pulse, while the receive pulse is still $u(t)$, {corresponding to a biorthogonality condition}. Furthermore, because the cross ambiguity function $\mathcal A_{u_{ce},u}(\cdot)$ is calculated between $u_{ce}(t)$ and $u(t- m\frac{T_0}{M})e^{j2\pi \frac{n}{NT_0}(t- m\frac{T_0}{M})}$, the problem to satisfy the orthogonality with respect to  $\Delta F=\frac{1}{NT_0}$ becomes how $u_{ce}(t)$ can have the specified periodicity within the range of $u(t-m\frac{T_0}{M})e^{j2\pi \frac{n}{NT_0}(t-m\frac{T_0}{M})}$ for $|m|\le M-1$.
We then have the following proposition:
\begin{prop}\label{l2}
  Let a pulse-train $u(t)$ be made up of $N$ square-root Nyquist pulses $a(t)$ with Nyquist interval $\frac{T_0}{M}$, which are temporally spaced by $T_0$. The resultant pulse train satisfies the biorthogonality property of
  \begin{equation}\label{afgceg}
    \mathcal A_{u_{ce},u}\left(m\frac{T_0}{M},n \frac{1}{NT_0}\right)= \delta(m)\delta(n),
  \end{equation}
  for $|m|\le M-1$ and $|n|\le N-1$, where
  $u_{ce}(t)$ is a {cyclically extended version of $u(t)$. Specifically, $u_{ce}(t)$ is a} periodic function with period $T_0$ during $-(M-1)\frac{T_0}{M}\le t \le (MN-1)\frac{T_0}{M}+T_a$.
\end{prop}
\begin{IEEEproof}
  See Appendix B.
\end{IEEEproof}
Note that the proof of Proposition \ref{l2} does not depend on $T_a$,
which indicates that the duration constraint of $a(t)$ in $u(t)$ can be removed. Once the appropriate CP and CS are added in accordance with (\ref{ncpcs}) (see Appendix B), where the extension parameter for CP and CS is $D=\lceil T_{a}/T_0 \rceil=\lceil 2Q/M \rceil$, the desired local/sufficient biorthogonality can be achieved as well.

\begin{figure*}
  \centering
  \includegraphics[width=17cm]{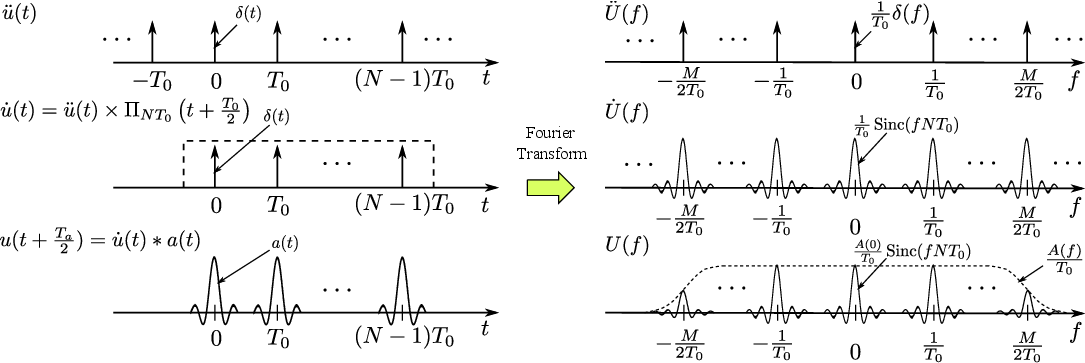}
  \caption{Derivation of $U(f)$}
  \label{derivation_uf}
\end{figure*}

\section{Important Properties of ODDM Modulation}
Given the general DDOP design, the transmit pulse of ODDM modulation becomes $u_{ce}(t)$.  When $M\gg 2Q$, we have  $2Q/M\approx 0$. Then, as proved in \cite{oddmicc22}, ODDM can employ the DDOP $u(t)$ without cyclic extension ($D=0$).

\subsection{TF Signal Localization}
The TF signal localization plays an important role in the analysis of modulation waveforms.
For example, the TF signal localization of CP-OFDM given in Fig. \ref{tf_lattice} explains its orthogonality with respect to the coarse JTFR in the conventional TF domain.
To understand the TF signal localization of an MC modulation, we need both the TD and FD representations of its transmit pulse.
In the following, we will derive $U(f)$, namely the FD representation of $u(t)$ corresponding to the case of $D=0$, as we usually have $M\gg 2Q$ in practice.

\subsubsection{Frequency Domain Representation of DDOP}

It is widely exploited that the FD representation of an impulse train
\begin{align}
  \ddot u(t)=\sum_{n=-\infty}^{\infty} \delta (t-nT_0),
\end{align}
is a Fourier series, which can also be written as an impulse train in the FD
\begin{align}
  \ddot U(f)=\frac{1}{T_0}\sum_{m=-\infty}^{\infty} \delta (f-\frac{m}{T_0}).
\end{align}
It is interesting to observe that the DDOP can be obtained from $\ddot u(t)$ by applying a rectangular window $\rect_{NT_0}\left(t+\frac{T_0}{2}\right)$ followed by {a filter with the impulse response $a(t)$}. Then, we have
\begin{align}
  u\left(t+\frac{T_a}{2}\right)=\dot u(t)*a(t),
\end{align}
where $\dot u(t)= \ddot u(t)\times \rect_{NT_0}\left(t+\frac{T_0}{2}\right)$ and $*$ {denotes convolution}. Since the multiplication and convolution in the TD correspond to the convolution and multiplication in the FD, respectively, we have
\begin{align}
  U(f) & =e^{-j2\pi f \frac{T_a}{2}} A(f)\dot U(f) \nonumber                                                                              \\
       & =e^{-j2\pi f \frac{T_a}{2}} A(f) \left(\ddot U(f)* e^{-j2\pi f \frac{(N-1)T_0}{2}}\sinc(fNT_0)\right), \nonumber                 \\
       & = \frac{e^{-j2\pi f \widetilde T}}{T_0} A(f)  \sum_{m=-\infty}^{\infty} e^{j2\pi \frac{m(N-1)}{2}}\sinc(fNT_0-mN) , \label{ufeq}
\end{align}
where $\dot U(f)= \ddot U(f)* e^{-j2\pi f \frac{(N-1)T_0}{2}}\sinc(fNT_0)$, $\widetilde T= (T_a+(N-1)T_0)/2$ and $A(f)$ is the Fourier transform of $a(t)$. It is noteworthy that the order of aforementioned windowing and filtering operations can be exchanged to obtain $u(t)$, leading to another representation of $U(f)$\cite{primer}.

Without loss of generality, let $M$ be an even number. Then, the derivation of  $U(f)$ {is graphically illustrated} in Fig. \ref{derivation_uf}, where the phase terms are omitted and the shapes of $\sinc(fNT_0-mN)$ are truncated for the purpose of display\cite{ddop,spcctalk}. Furthermore, it is assumed in Fig. \ref{derivation_uf} that $a(t)$ has a small roll-off factor and therefore $A(f)$ has a steep edge decay to contain $(M+1)$ Sinc functions centered at periods of $\frac{1}{T_0}$ from $-\frac{M}{2T_0}$ to $\frac{M}{2T_0}$. Now, it becomes clear that $\sinc(fNT_0-mN)$ and  $A(f)$ correspond to the orthogonality with respect to $\Delta F=\frac{1}{NT_0}$ and $\Delta T= \frac{T_0}{M}$, respectively.

The operations and results in Fig. \ref{derivation_uf} can be extended straightforwardly to \emph{other pulse-trains with different subpulses $a(t)$} including apparently the Nyqyist pulse with Nyquist interval $\frac{T_0}{M}$. In fact, we can choose the subpulse to determine the \emph{envelope} of the FD representation of the pulse-train, which corresponds to the {bandwidth of the pulse-train}. Meanwhile, we are also free to choose $N$ and $T_0$ for beneficially manipulating the FD representation of the pulse train under the envelope, which is $\dot U(f)$.

\subsubsection{DDOP as Virtual 2D Pulse}\label{2dpulse}

Being a continuous-time function, a pulse can be described either by its TD representation $g(t)$ or FD representation $G(f)$, where these two representations are \emph{tightly} bound by the (inverse) Fourier transform and therefore dependent on each other. This dependency is exactly the reason why the {TFA of the pulse} has a lower bound (Gabor limit) of $1/(4\pi)$ corresponding to the Heisenberg uncertainty principle. As a result, although we may be able to present {the TF localization of a pulse} in the TF domain by illustrating its time and frequency representations together as shown in Fig. \ref{tf_lattice}, the time and frequency variables of the pulse's TF domain are inter-dependent. Therefore, they  cannot form a \emph{real} 2D domain $(t,f)$ to let us design a 2D pulse/filter that can be denoted by $\mathsf g(t,f)$. {In other words, a pulse is always a 1D function.}

On the other hand, we would point out that the delay variable $\tau$ and the Doppler variable $\nu$, namely the time and frequency variables of the channel's TF domain, are independent. Therefore, they can form a real 2D domain, and we do have 2D impulse responses for LTV channels, represented by the spreading function $h(\tau, \nu)$.

If we can artificially introduce an \emph{extra} time variable, a pulse having a pulse-train structure may be considered as a ``virtual" 2D pulse to match the signal to the channel.
For example, the intervals between the subpulses and the time during an interval may be considered as two potential variables of the pulse-train or the virtual 2D pulse. This is similar to \emph{slow time} and \emph{fast time}, which have been widely used in radar waveform design\cite{radar_handbook,radar_sp}.
In particular, for the pulse-train $u(t)$, although its subpulse $a(t)$ has a tightly bound pair of time variable $t$ and frequency variable $f$, we can repeat the subpulse $a(t)$ and introduce an extra time variable $\dot t$, {the minimum unit of which} is $T_0$. 
By letting $t=\tau+\dot t$, we can virtually represent $u(t)$ as a 2D function $\mathsf u(\tau, \dot t)$ subject to the following constraints
\begin{align}\label{2dcon}
  \begin{cases}
    \tau \in [0,T_0),                  \\
    \dot t= n T_0, n = 0, \cdots, N-1, \\
    \mathsf u(\tau, \dot t)=\mathsf u(\tau, 0).
  \end{cases}
\end{align}
Note that $\dot t$ is independent of $\tau$ when  (\ref{2dcon}) holds.
Now we may be able to apply the Fourier transform to $\dot t$ and obtain a ``virtual" 2D domain $(\tau, \nu)$.
Thus, $u(t)$ may be represented using a virtual 2D pulse $\mathsf u(\tau, \dot{t})$ characterized by a pair of two independent parameters $\tau $ and \(\dot{t}\) or $\tau $ and $ \nu$.
The last two constraints in (\ref{2dcon}) actually correspond to $\dot U(f)$, the FD representation of the pulse train under the envelope we mentioned before.

\begin{figure}
  \centering
  \includegraphics[width=8cm]{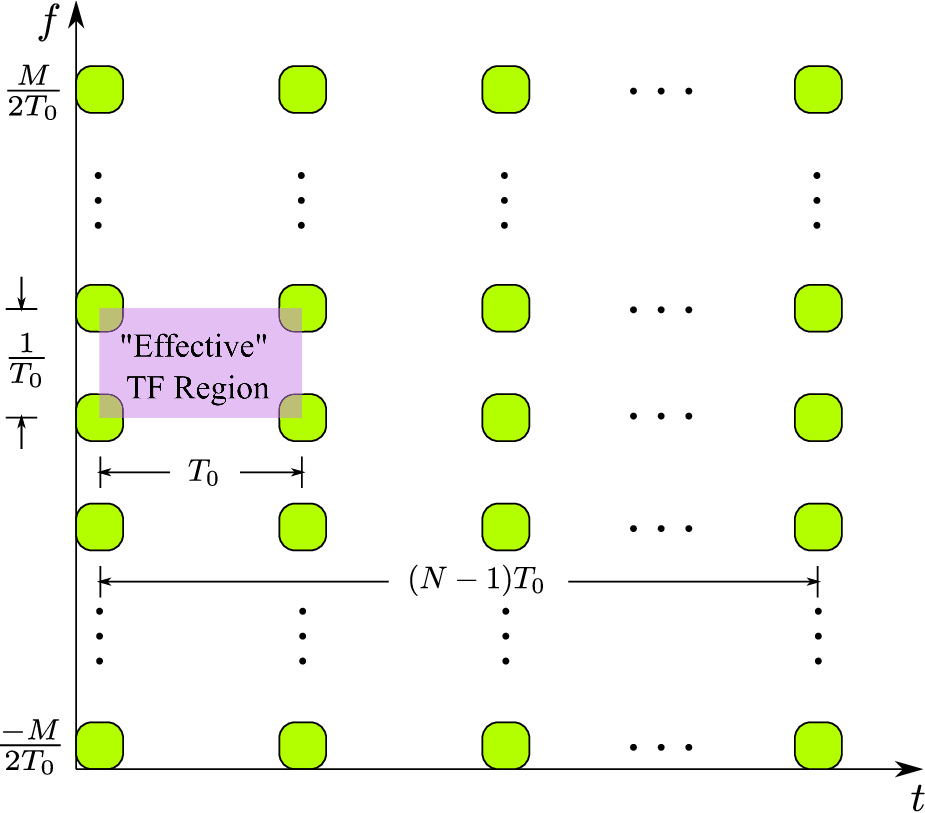}
  \caption{Simplified TF signal localization of DDOP}
  \label{tfl_ddop}
\end{figure}

\begin{figure*}
  \centering
  \includegraphics[width=\linewidth]{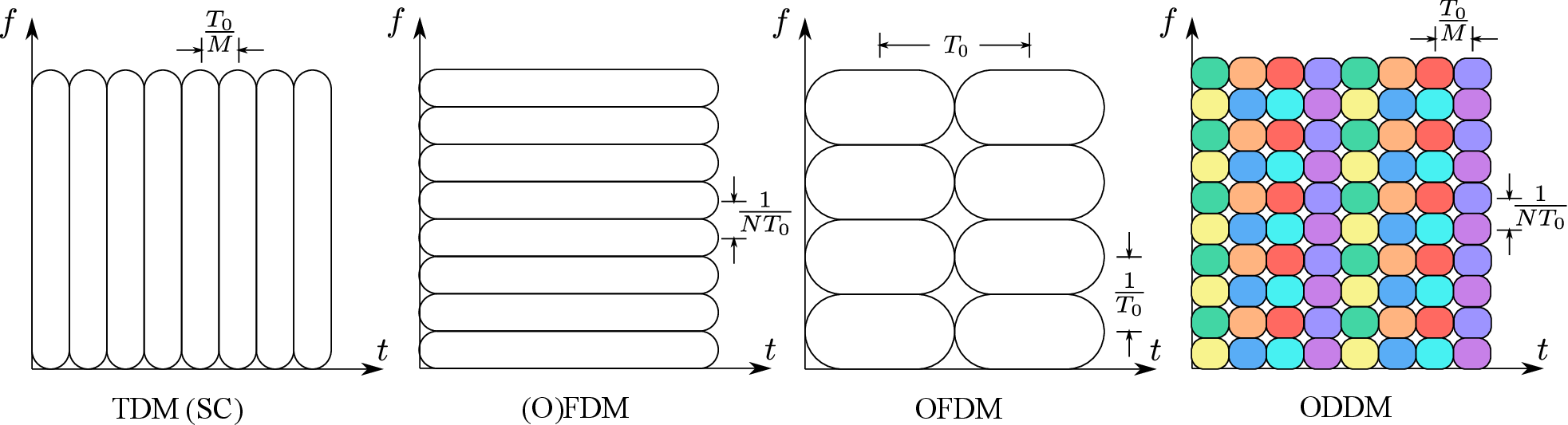}
  \caption{Comparison of simplified TF signal localization, $M=4$ and $N=2$.}
  \label{cms}
\end{figure*}

It should be noted that introducing $\dot t$ does not mean that we can escape from the tight relation between the unconstrained $t$ and $f$ to create a new dimension for pulse design. As we can see from $u(t)$ and $U(f)$, it just causes the signal to be {equally-spaced} distributed in both the TD and FD, and partitions the TF region of the signal into $MN$ small ``effective" regions.
Inside an effective TF region, the orthogonality with respect to the delay resolution $\frac{T_0}{M}$ and the Doppler resolution $\frac{1}{NT_0}$ is achieved by $a(t)$ in $u(t)$ and $\sinc(fNT_0)$ in $U(f)$, respectively. With the aid of $u(t)$ and $U(f)$, the simplified TF signal localization of DDOP is illustrated in Fig. \ref{tfl_ddop}.

\subsubsection{Comparison of TF Signal Localization}
To transmit $MN$ QAM symbols, an MC modulation scheme employs $MN$ orthogonal pulses corresponding to its JTFR and results in its own TF signal localization. {Based on Fig. \ref{tfl_ddop}}, the comparison between ODDM and other modulation schemes in terms of their \emph{simplified} TF signal localization can be schematically illustrated in Fig. \ref{cms}\footnote{From (\ref{xtotfs}) and (\ref{isfft}), we can readily see that OTFS is a precoded OFDM. Each precoded symbol occupies the same time-frequency resource as OFDM. We have thus omitted OTFS in Fig. \ref{cms}.}, where $M=4$, $N=2$
.
It should be noted that the bandwidth and
duration of each pulse in these modulation schemes are not explicitly presented in Fig. \ref{cms}, and for ODDM, $M$ has to be a large enough integer to have a reasonable extension parameter $D=\lceil 2Q/M \rceil$ for the general DDOP.
From Fig. \ref{cms}, one can observe that:
\renewcommand{\labelenumi}{\arabic{enumi})}
\begin{enumerate}
  \item For SC modulation, which is a \ac{tdm} scheme, the $MN$ QAM symbols are conveyed by $MN$ square-root Nyquist pulses for the Nyquist interval  $\frac{T_0}{M}$. The pulses are orthogonally overlapped in the TD.
  \item For a frequency-division multiplexing (FDM) scheme, such as for example OFDM associated  with frequency resolution of $\frac{1}{NT_0}$, $MN$ QAM symbols are conveyed by $MN$ rectangular pulses $\rect_{NT_0}(t)$ modulated by $MN$ subcarriers, respectively. The pulses are inseparably overlapped in the TD however they are orthogonally overlapped in the FD.
  \item For conventional OFDM having a frequency resolution of $\frac{1}{T_0}$ and time resolution $T_0$, $MN$ QAM symbols are conveyed by $N$ OFDM symbols, where each OFDM symbol has $M$ rectangular pulses $\rect_{T_0}(t)$ modulated by $M$ subcarriers, respectively. Since $N$ OFDM symbols are isolated in the TD, the inter-symbol pulses {are not overlapped either in the TD or in the FD}, while the intra-symbol pulses are inseparably and orthogonally overlapped in the TD and FD, respectively.
  \item For ODDM having a frequency resolution of $\frac{1}{NT_0}$ and a time resolution of $\frac{T_0}{M}$, $MN$ QAM symbols are conveyed by $M$ pulse trains $u(t)$ modulated by $N$ subcarriers, respectively. These pulses are overlapped orthogonally in both the TD and FD.
\end{enumerate}

Since the overlapping of pulses is the key to high bandwidth efficiency\cite{ofdm_oqam_chang_66,tlomcm_95}, it is meaningful to investigate the bandwidth efficiency of the ODDM modulation, the pulses of which are overlapped in both the TD and FD.
Recall that for the TF region is bounded by $B_x$ and $T_x$,
we have a DoF around $B_xT_x$. Then, for a given $M$ and $N$, we can calculate the necessary $B_x$ and $T_x$ for each modulation scheme and obtain their bandwidth efficiencies accordingly.

\subsection{Bandwidth Efficiency}
Let the square-root Nyquist pulse be a \ac{rrc} pulse\footnote{Due to its good TF localization, the RRC pulses modulated by spreading code/sequence were adopted by \ac{3g} \ac{cdma} communications.} with roll-off factor  $\rho$. Upon recalling that $T_a=2Q\frac{T_0}{M}$, we have
\begin{align}
  \eta_{\textrm{TDM}} & =\frac{MN}{\frac{M}{T_0}(1+\rho)((MN-1)\frac{T_0}{M}+T_a)}\nonumber \\
                      & =\frac{1}{(1+\rho)(1+\frac{2Q-1}{MN}) }.
\end{align}
For (O)FDM, because $|\sinc(fNT_0)|$ decays as $1/f$, it can be treated as negligibly small {beyond the $\mathcal K$th zero-crossing on both sides of the main lobe}. In other words, the bandwidth of $g(t)=\rect_{NT_0}(t)$ is considered as $B_g=2\mathcal K\frac{1}{NT_0}$. Then, we have
\begin{align}
  \eta_{\textrm{(O)FDM}}=\frac{MN}{(\frac{MN-1}{NT_0}+\frac{2\mathcal K}{NT_0})NT_0}=\frac{1}{1+\frac{2\mathcal K-1}{MN}}.
\end{align}

Let us now consider CP-OFDM. Because of the delay spread of the channel, $g(t)$ is $\rect_{T_0+T_{cp}}(t)$ and $T_{cp}=L\frac{T_0}{M}$. Therefore, we have
\begin{align}
  \eta_{\textrm{CP-OFDM}} & =\frac{MN}{(\frac{M-1}{T_0}+\frac{2\mathcal K}{T_0+T_{cp}})NT_0(1+\frac{L}{M})},\nonumber \\
                          & =\frac{1}{(1+\frac{M(2\mathcal K-1)-L}{M(M+L)})(1+\frac{L}{M})}.
\end{align}
For ODDM, because the cyclic extension of $u(t)$ is equivalent to the frame based CP and CS, we have

\small
\begin{align}
  \eta_{\textrm{ODDM}} & =\frac{MN}{\frac{MN(1+\rho)+N-1}{NT_0}((N-1+2D)T_0+\frac{(M-1+L)T_0}{M}+T_a)}, \nonumber \\
                       & =\frac{1}{(1+\rho+\frac{N-1}{MN})(1+\frac{2DM+L+2Q-1}{MN})}.
\end{align}
When $2Q\ll M$, we can let $D=0$. Then, the ODDM frame only has a single CP corresponding to the delay spread of the channel. For such an ODDM frame with CP only (CP-ODDM), we have
\begin{align}
  \eta_{\textrm{CP-ODDM}}
  =\frac{1}{(1+\rho+\frac{N-1}{MN})(1+\frac{L+2Q-1}{MN})}.
\end{align}

\normalsize

Considering $\frac{1}{T_0}=15$kHz and the \ac{eva} channel model of \cite{eva_channel_model} associated with delay spread $2510$ns, a bandwidth efficiency comparison of these modulation schemes is shown in Fig. \ref{secomp}, where we have $M=512$, $L=20$, and $\mathcal K=11$ corresponds to the $99$\% fractional power containment bandwidth. Note that due to the duality between time and frequency, $Q=11$ corresponds to a $99$\% power containment duration when $\rho=0$. Furthermore, for the same level of power containment, $Q$ decreases as $\rho$ increases, and $Q>11$ is chosen for the comparison shown in Fig. \ref{secomp}. One can see that ODDM has similar bandwidth efficiency to SC modulation. {For a moderately large $N$ and an appropriate $\rho$}, ODDM has better bandwidth efficiency than CP-OFDM, while the (O)FDM with $\Delta F=\frac{1}{NT_0}$ has the highest bandwidth efficiency. Note that the (O)FDM associated with $\Delta F=\frac{1}{NT_0}$ has $MN$ subcarriers and therefore it is extremely ``expensive" to implement. It is also noteworthy that a more accurate comparison can be performed using the OOBE-based approach of \cite{hama}.

\begin{figure}
  \centering
  \includegraphics[width=8.5cm]{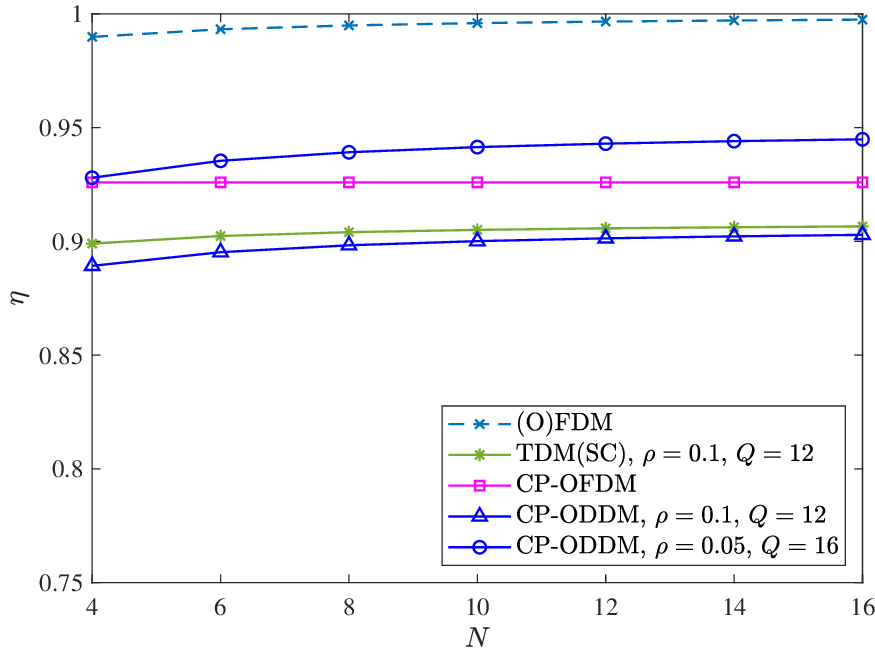}
  \caption{bandwidth efficiency of modulation schemes, $L=20$.}
  \label{secomp}
\end{figure}

\begin{figure*}
  \centering
  \includegraphics[width=0.96\linewidth]{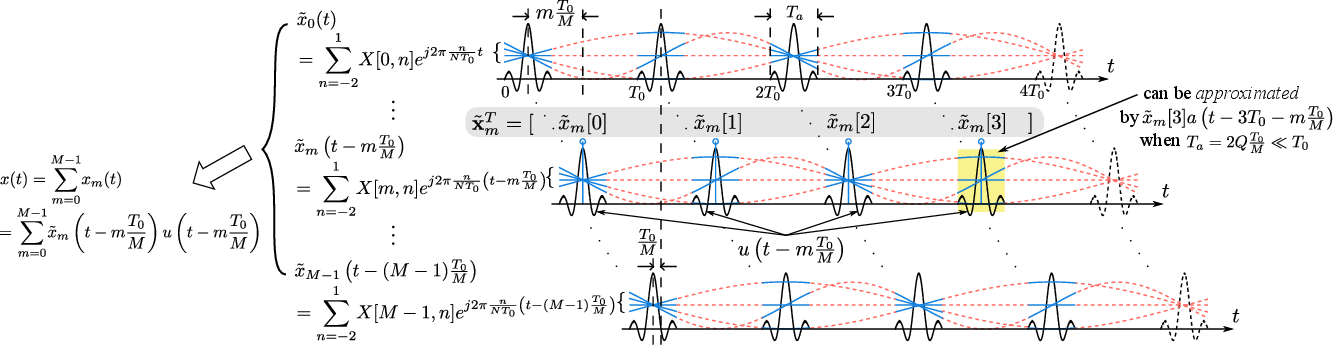}
  \caption{ODDM waveform without CP and CS, when $N=4$.}
  \label{oddmwaveform}
\end{figure*}

\subsection{Implementation Methods}
Being a standard MC modulation, ODDM can be implemented straightforwardly via the analog and digital approaches mentioned in Section \ref{tfmc_imple_method}. In particular, we can substitute $\Delta T=\frac{T_0}{M}$ and  $\Delta F=\frac{1}{NT_0}$ or $T=NT_0$ into Figs. \ref{adi1}-\ref{adi2n_ifft3}, and replace $g(t)$ by $u(t)$. However, due to the long duration of $u(t)$, these direct implementations have a high complexity, even if we generate $\tilde x_m(t)$ using the IFFT.

Figure \ref{oddmwaveform} shows an ODDM waveform without CP and CS, when $N=4$. One can see that due to the limited duration of $a(t)$, $x_m(t)$ namely $\tilde x _m(t-m\frac{T_0}{M})$ shaped (multiplied) by $u(t-m\frac{T_0}{M})$ becomes discontinuous with $N$ segments, {each with a length} $T_a$. When $T_a\ll T_0$, it has been proved in {\cite[Appendix A]{oddm}} that instead of {using} $u(t-m\frac{T_0}{M})$-based pulse-shaping, we can generate the discrete samples $\tilde {\mathbf x}_m^T$ and then filter them with $a(t)$ to \emph{approximate} $x_m(t)$. For example, the fourth segment of $x_m(t)$ can be approximately generated by filtering or sample-wise pulse-shaping $\tilde x_m[3]$ with $a(t)$, as shown in Fig. {\ref{oddmwaveform}. This approximation is actually \emph{valid for any other subpulses}, as long as $T_a\ll T_0$. As a result, we have ODDM approximated by a filtered OFDM, where the filter is a \emph{wideband} filter to retain the frequency diversity created by the sampling-induced aliasing, see detailed explanations in {Remarks 2 and 4} of \cite{oddm}.

\begin{figure}
  \centering
  \includegraphics[width=7cm]{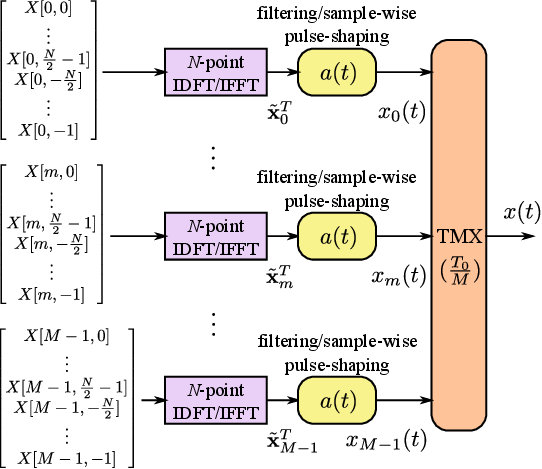}
  \caption{\emph{Approximate} implementation of ODDM}
  \label{oddm_di1}
\end{figure}

\begin{figure}
  \centering
  \includegraphics[width=8cm]{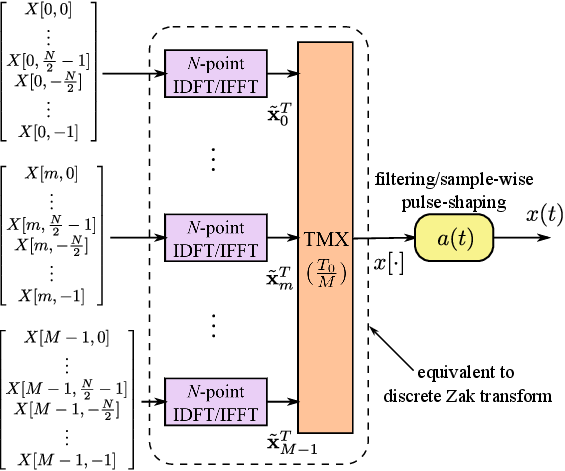}
  \caption{Simplified \emph{approximate} implementation of ODDM}
  \label{oddm_di2}
\end{figure}

The wideband filtered OFDM based approximation leads to the low-complexity implementation of ODDM shown in Fig. \ref{oddm_di1}, where a filtering or sample-wise pulse-shaping is employed \cite{oddm}. Moreover, because the $M$ branches of Fig. \ref{oddm_di1} share the same filter $a(t)$,
we can exchange the order of the TMX and the filters to further simplify the implementation. The resultant simplified approximate implementation of ODDM using a digital TMX and a single wideband filter is shown in Fig. \ref{oddm_di2}. The generated \emph{approximate} ODDM waveform can be written as \cite{2024_TCOM_JunTong_ODDMoverPhysicalChannels}
\begin{align}\label{xtoddmapprox}
  \dot x(t) =
  \sum_{m=0}^{M-1} \sum_{n=-N/2}^{N/2-1} X[m, n]  \dot u_{m,n}(t),
\end{align}
where
\begin{align}\label{gammat}
  \dot u_{m,n}(t) = \sum_{\dot{n}=0}^{N-1} e^{j 2 \pi \frac{n \dot{n} }{N}} a\left(t-\dot{n} T_0 -m \frac{T_0}{M}  \right).
\end{align}

It is noteworthy that the combination of {$M$ branches of $N$-point IDFT/IFFT having $T_0$-interval output samples} and the $\frac{T_0}{M}$-interval digital TMX is equivalent to a discrete Zak transform\cite{zak}. Because a discrete-time  OTFS sequence can be generated using the discrete Zak transform \cite{hadaniyt}, as indicated in Remark 4 of \cite{oddm}, a \emph{digital} namely discrete-time OTFS signal filtered or sample-wise pulse-shaped by $a(t)$ approximates the ODDM waveform.
Meanwhile, it should be noted that the wideband filter $a(t)$ can also be implemented digitally, followed by a DAC having a high enough sampling rate.


\subsection{Potentials for ISAC}
At the time of writing, {ISAC} is regarded as a promising technology for next-generation wireless communications to intelligently utilize the precious spectrum. An ISAC system is essentially a dual-functional radar-communication (DFRC) system, where the backscattered signals are used for estimating object or {scattering} parameters, such as range and velocity corresponding to delay and Doppler, respectively. Because radar sensing and communication functions have different sometimes even \emph{opposing} requirements on waveform properties, the main challenge in the ISAC system development is to design a suitable waveform that can simultaneously perform these two tasks well\cite{sturm2011}.

A waveform is characterized by its format and diverse other parameters including bandwidth, duration, TF resolutions, etc. Regarding the format,
because the conventional radar waveforms, such as frequency modulated continuous waveforms and chirp signals, only have limited communication capability\cite{ma2020}, the communication waveforms including the SC and MC modulations become the primary choice for DFRC systems\cite{sturm2011,paul2017,ma2020}. Regarding the parameters, from a radar sensing perspective, the waveform is expected to {be wideband with a long duration, which corresponds} to both high delay and Doppler resolutions and subsequently  good sensing performance. On the other hand, from the communications perspective, a high throughput and low latency require a wideband waveform having a short duration, while a narrowband signal with a relatively long duration can ease the channel equalization.
Therefore, first we may have to determine the {bandwidth and duration of the ISAC waveform}, according to the {radar sensing and communication applications under consideration}.

Once the bandwidth and duration constraints are given, we can design a communication signal based ISAC waveform, by taking into account the performance metrics of both functions. For communications, the metrics include the achievable rate, the bandwidth efficiency, the equalization complexity, etc.
For sensing, considering the classic correlation-based approaches, popular metrics include different characteristics of the TD autocorrelation function, for example, the main-lobe width, the peak to side-lobe level, and the \ac{isll} \cite{guo_radarcon_2014}. Without exploiting the {information content of the signal}, the correlation based sensing approaches have limited performance, especially in the presence of large Doppler shifts\cite{sturm2011}. In particular, for the SC waveform (in combination with spread-spectrum techniques) designed for optimizing the time-domain autocorrelation, the estimation of Doppler/velocity is difficult\cite{sturm2011}. Meanwhile, by explicitly exploiting the {information content of the signals in MC modulations}, {radar sensings} in the ``Modulation Symbol" domain \cite{berger_jstsp_2010,sturm2011} can achieve superior performance over the correlation based approaches.

It should be noted that the radar sensing is exactly constituted by the estimation of the backscattered channel\cite{paul2017}, which is also an ESDD channel.
The rationale behind the ``Modulation Symbol" domain based sensing approaches is simply that the transmit information-bearing symbols are known at the radar receiver and therefore can be used as pilots to perform pilot-based channel estimation. By contrast, the correlation based sensing can be viewed as a blind channel estimation, which usually has inferior performance.

Recall that ODDM is an impulse-function-based transmission technique designed for ESDD channels. The estimation of the forward communication ESDD channel can be performed straightforwardly with the aid of DD domain pilots. On the other hand, radar sensing or the estimation of the backscatter ESDD channel only requires an appropriately extended frame based CP corresponding to the longer delay spread of the backscatter ESDD channel.
Meanwhile, because the ESDD channel estimation is a necessary part of an ODDM receiver, an ODDM system can be interpreted as an ISAC system, where the communication and radar sensing have been seamlessly integrated.

An interesting interpretation for the ISAC capability of the ODDM waveform can be obtained from the characteristics of the {ambiguity function of DDOP}.
Notice that the TD autocorrelation function is an ambiguity function without frequency shift. A more appropriate metric for radar sensing may be the normalized \ac{isll} of the ambiguity function defined as\cite{keskin_tsp_2021}
\begin{align}\label{isl}
  \textrm{ISLL}_u=\frac{\int\int_{R_s} |A_{u,u}(\tau, \nu)|^2d\tau d\nu}{| R_s||A_{u,u}(0, 0)|^2},
\end{align}
where $R_s$ denotes the side-lobe region in the DD domain.
Further considering the bandwidth and duration constraints and the corresponding  delay and Doppler resolutions, we can modify (\ref{isl}) to define a normalized \ac{sisll} of the ambiguity function as
\begin{align}\label{snisl}
  \textrm{SISLL}_u=\frac{\sum_{m=0}^{L-1}\sum_{n=-K}^{K}|A_{u,u}(\frac{m}{\mathbb W},\frac{n}{\mathbb T})|^2}{|A_{u,u}(0, 0)|^2},
\end{align}
{the minimum of which} can be achieved by the DDOP.

Meanwhile, because the repeatedly sent pulses in a pulse-Doppler radar can be viewed as a pulse-train\cite{radar_handbook}, one can see  that the DDOP in Fig. \ref{shape_ut} is exactly a kind of pulse-Doppler radar waveform. Note that the measurement of Doppler/velocity depends on the repeat subpulses or the signal structure of pulse-train. By using the root Nyquist pulse as  the subpulse in the pulse train, we can also benefit the measurement of delay/range in radar sensing function, as well as the alignment of interferences in communication function. As a result, the DDOP-based ODDM \emph{combines the key characteristics of radar and communication waveforms} and becomes a natural waveform choice for ISAC.

\begin{table}[t]
  \centering
  \caption{Simulation Parameters}
  \def\arraystretch{1.1}
  \begin{tabular}{|l|r|}
    \hline
    \thead[c]{Parameter}    & \thead[c]{Value} \\
    \hline
    Carrier frequency $f_c$ & 5 GHz            \\
    \hline
    $1/T_0$                 & 15 kHz           \\
    \hline
    $M$                     & 512              \\
    \hline
    CP length               & 3.125$\mu$s      \\
    \hline
    Modulation alphabet     & 4-QAM            \\
    \hline
    UE speed (km/h)         & 80, 120, 500     \\
    \hline
  \end{tabular}
  \label{tab:simpara}
\end{table}

\section{Numerical Results}
In this section, simulations are conducted to verify the performance of the ODDM modulation. The simulation parameters are shown in Table \ref{tab:simpara}. For the doubly-selective channel, we adopt the EVA model \cite{eva_channel_model}, where each path has a single Doppler generated using Jakes' formula $\nu_p = \nu_{\textrm{max}} \cos(\phi_p)$, the maximum Doppler $\nu_{\textrm{max}}$ is determined by the \ac{ue} speed and $\phi_p$ is uniformly distributed over $[-\pi,\pi]$. It is noteworthy that the EVA channel has not only off-grid channel taps on the delay axis, but also possible off-grid Dopplers. Also, a RRC pulse with a roll-off factor of $\rho$ and a duration of $2Q\frac{T_0}{M}$ is employed as $a(t)$.

The \ac{psd} comparison of the modulated signals with various RRC pulse parameters is shown in Fig. \ref{oddm_psd}. From this figure, we can see that the PSD of the proposed ODDM signals can maintain low OOBE.  In addition, we also see that by tuning the roll-off factor,
a trade-off between the excess bandwidth and OOBE can be
struck to achieve the desirable bandwidth efficiency.

\begin{figure}
  \centering
  \includegraphics[width=8.5cm]{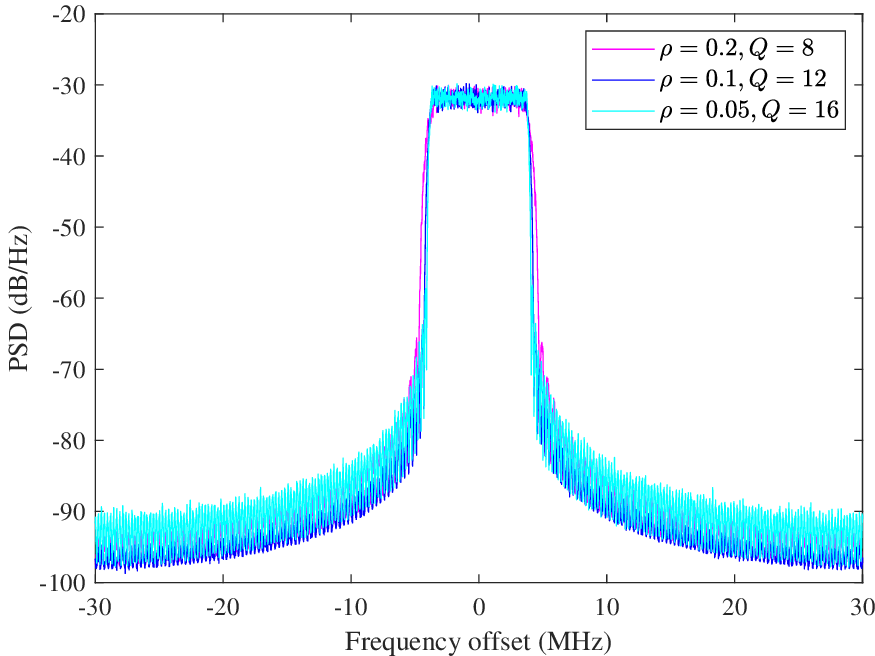}
  \caption{PSD, $M=512$, $N=32$, $4$-QAM.}
  \label{oddm_psd}
\end{figure}

The \ac{nmse} between the approximated ODDM signal and the exact ODDM waveform is defined as
\begin{align}
  \textrm{NMSE}_{\dot x}=\frac{\int |\dot x(t)-x(t)|^2 dt}{\int |x(t)|^2 dt},
\end{align}
where $x(t)$ is the exact ODDM waveform given in (\ref{xtoddm}) and $\dot x(t)$ is the approximated ODDM signal given in (\ref{xtoddmapprox}).
The NMSE results for various roll-off factor $\rho$ and duration of pulses is shown in Fig. \ref{xt_appro_nmse}.  We can see from this figure that the NMSE is not significantly affected by the parameter $Q$ of the pulse, and the NMSE decreases as the roll-off factor of the pulse employed increases. The figure also demonstrates that the simplified approximate implementations in Fig. \ref{oddm_di1} and Fig.
\ref{oddm_di2} can generate very close ODDM waveform as the NMSE between them is below $-40$dB.

\begin{figure}
  \centering
  \includegraphics[width=8.5cm]{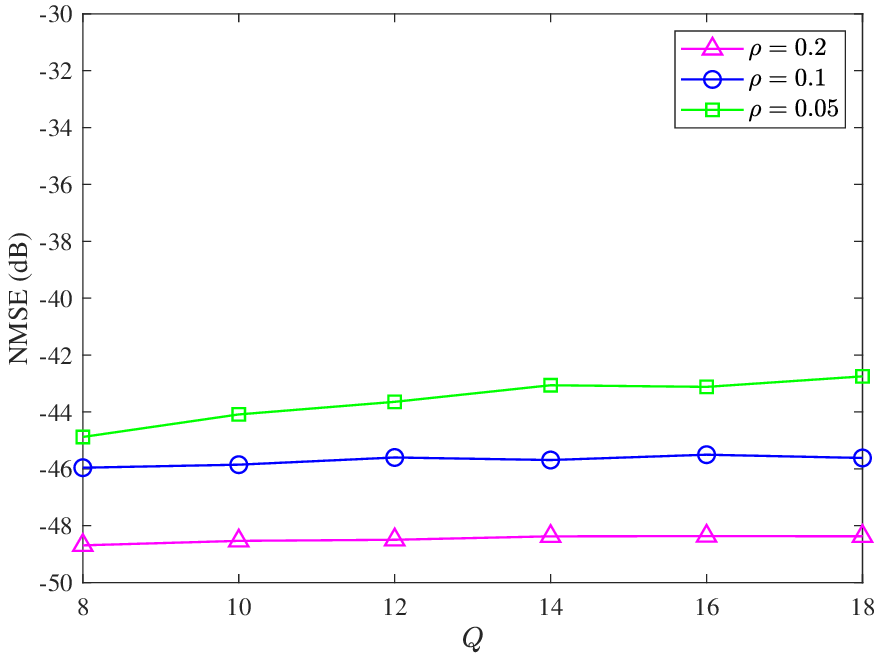}
  \caption{NMSE of approximated ODDM waveform, $M=512$, $N=32$, $4$-QAM.}
  \label{xt_appro_nmse}
\end{figure}

We now evaluate the BER performance of the uncoded ODDM modulation.\footnote{Novel code designs for ODDM is a crucial research problem. A code design that specifically focuses on ODDM was recently proposed in \cite{2025_TCOM_Oliver}, while the performance of ODDM with convolutional codes has been presented in \cite{oddm} (see Figs. 9 and 10).}
The signal detection is based on the message passing algorithm \cite{viterbo_twc_2018} and the DD domain channel matrix $\mathbf H$ in (\ref{H}).
Fig. \ref{ber_N32_S500_rho} shows the BER of the ODDM signals with $M=512$, $N=32$ and 4-QAM. In the simulation, the maximum UE speed is $500$km/h and the roll-off factor of the pulse is chosen as $0.05$, $0.1$ and $0.2$. This figure demonstrates that the ODDM signal achieves almost the same BER for various roll-off factor of the pulse.

\begin{figure}
  \centering
  \includegraphics[width=8.5cm]{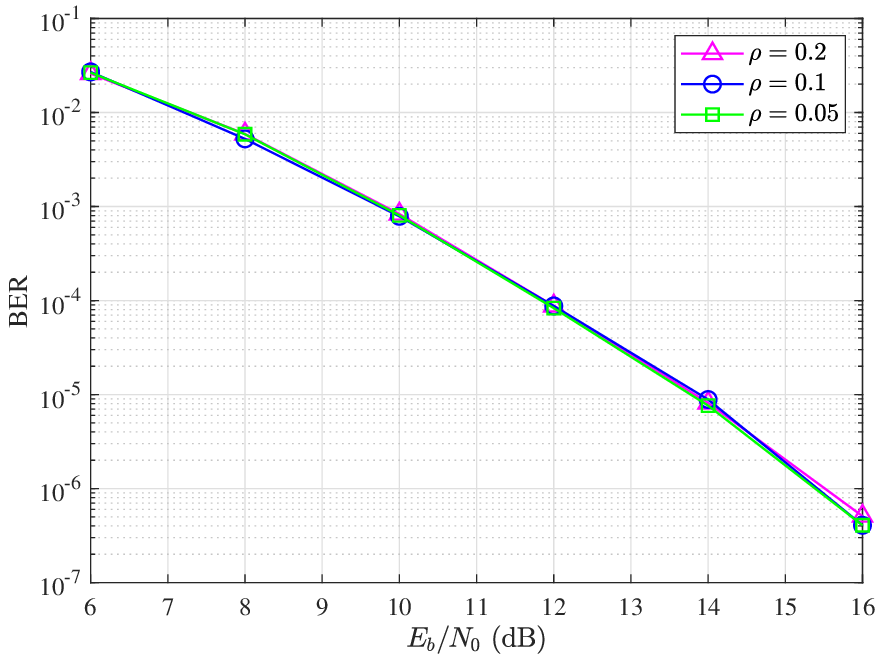}
  \caption{BER comparison, $M=512$, $N=32$, $4$-QAM, $500$km/h.}
  \label{ber_N32_S500_rho}
\end{figure}

\begin{figure}
  \centering
  \includegraphics[width=8.5cm]{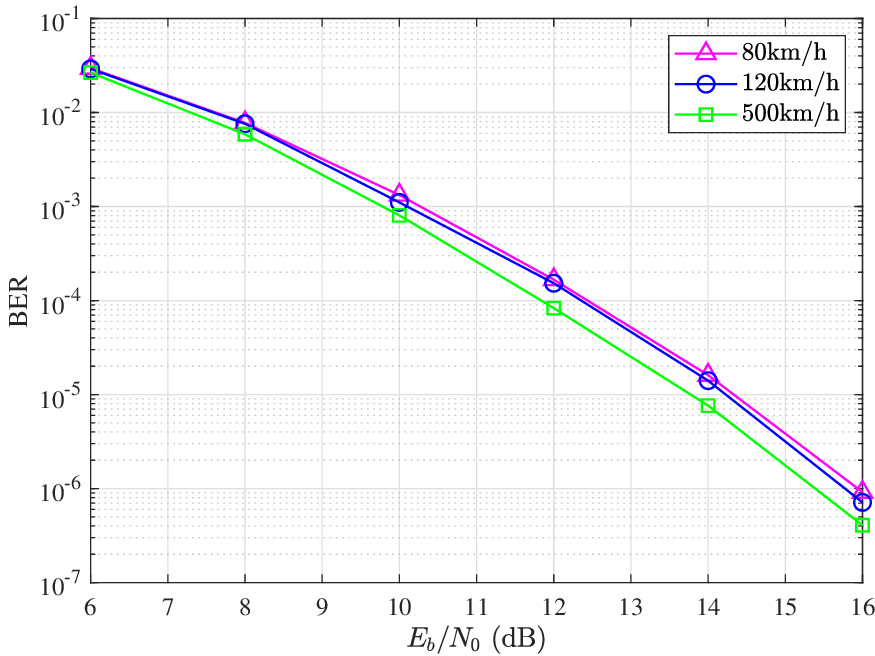}
  \caption{BER comparison, $M=512$, $N=32$, $4$-QAM, $\rho=0.05$, $Q=16$.}
  \label{ber_N32_rho01_q12_S}
\end{figure}

\begin{figure}
  \centering
  \includegraphics[width=8.5cm]{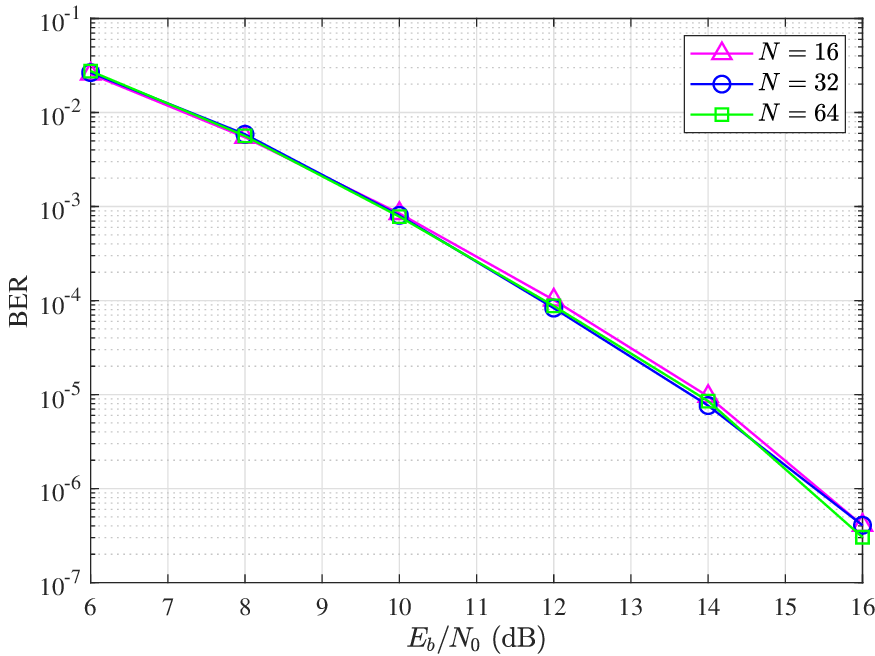}
  \caption{BER comparison, $M=512$, $4$-QAM, $\rho=0.05$, $Q=16$, $500$km/h.}
  \label{ber_S500_rho01_q12_N}
\end{figure}

Fig. \ref{ber_N32_rho01_q12_S}  shows the BER of the proposed ODDM system
with $4$-QAM signals, $M= 512$, $N= 32$, and UE speed of $80$km/h, $120$km/h and $500$km/h.
The figure demonstrates that the ODDM signals achieve almost the same BER performance over the high-mobility channels regardless of the UE speed, which means that ODDM signals are robust against Doppler shifts.
Meanwhile, Fig. \ref{ber_S500_rho01_q12_N}  illustrates the BER of the proposed ODDM system
with $4$-QAM signals, $M= 512$, $N= \{16, 32, 64\}$ and UE speed of $500$km/h.
  {The figure shows that the BER performance of the ODDM signals also remains almost the same for various values of $N$.}

A three-dimensional plot of the ambiguity function in (\ref{afgceg}), a.k.a ambiguity surface\cite{ofdmvsfbmc}, is shown in Fig. \ref{as}, where $\Delta F=\frac{1}{NT_0}$, $\Delta T=\frac{T_0}{M}$ with $M=32$, $N=8$, and the RRC pulse $a(t)$ has the roll-off factor $\rho=0.1$ and $Q=20$. Because for this parameter setting, the extension parameter for CP and CS is $D=2$, and we adopt the general DDOP design. The corresponding 2D plots of $\left|\mathcal A_{u_{ce},u}\left(\tau,\nu\right)\right|$ with $\nu=0$ and $\tau=0$ are also given in Figs. \ref{as_n0} and \ref{as_m0}, respectively. One can see that with appropriate CP and CS, the DDOP can achieve the sufficient orthogonality within $|m|\le M-1$ and $|n|\le N-1$. For $|m|\ge M$ or $|n|\ge N$, the ambiguity function repeats with time period $T_0$ and frequency period $\frac{1}{T_0}$, if we further extend
the CP and CS. These figures indicate the great potential of the DDOP-based ODDM signals for ISAC applications.

\begin{figure}
  \centering
  \includegraphics[width=8.8cm]{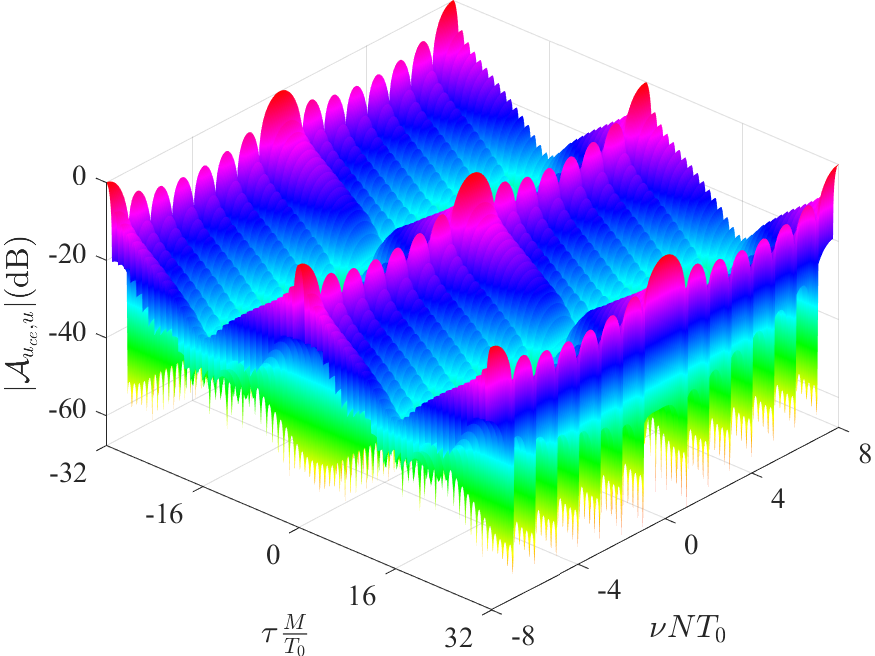}
  \caption{$\left|\mathcal A_{u_{ce},u}\left(\tau,\nu\right)\right|$ for $M=32$ and $N=8$.}
  \label{as}

\end{figure}

\begin{figure}
  \centering
  \includegraphics[width=8.5cm]{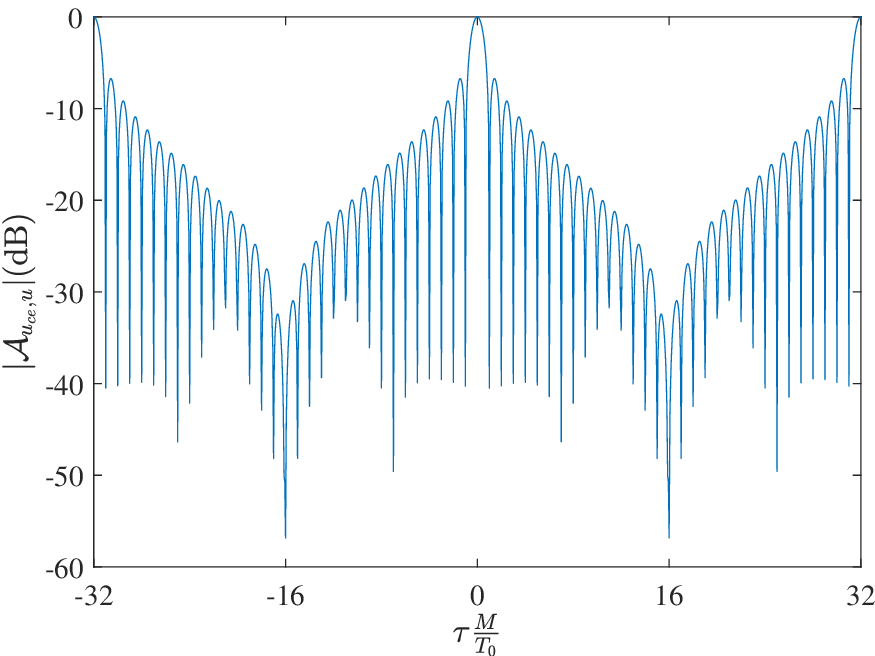}
  \caption{$\left|\mathcal A_{u_{ce},u}\left(\tau,0\right)\right|$ for $M=32$ and $N=8$.}
  \label{as_n0}
\end{figure}

\begin{figure}
  \centering
  \includegraphics[width=8.5cm]{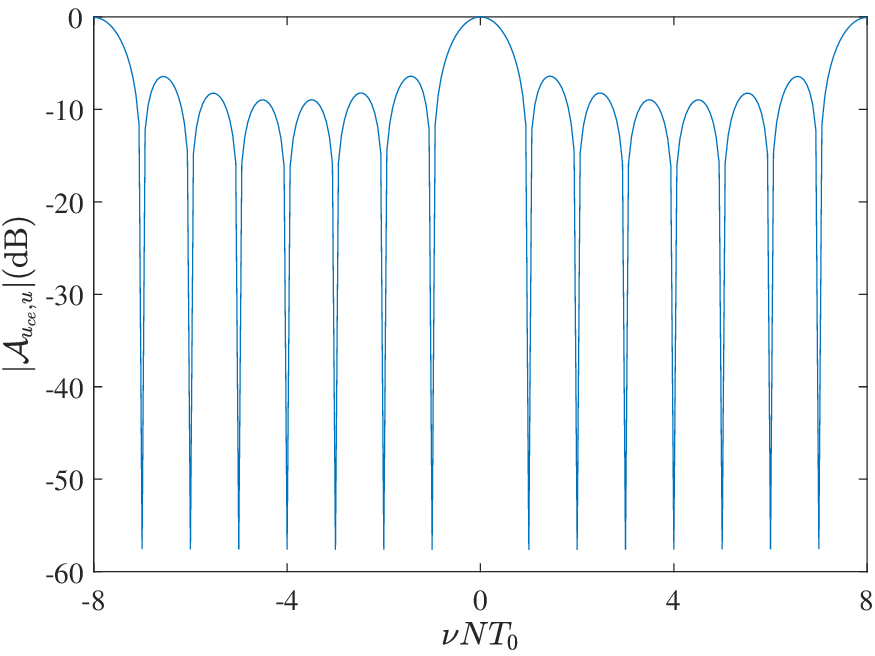}
  \caption{$\left|\mathcal A_{u_{ce},u}\left(0,\nu\right)\right|$ for $M=32$ and $N=8$.}
  \label{as_m0}
\end{figure}

\section{Conclusions and Future Research}
An in-depth look into DDOP and the corresponding ODDM modulation has been presented to unveil their unique characteristics.
We first revisit the conventional \ac{tfmc} modulation schemes in terms of their transmission strategy, the channel-oriented orthogonal or biorthogonal pulses and the resultant bandwidth efficiency. Next, we address the 
time-varying property of the DD domain channel's 2D impulse response and clarify the unique and innovative transmission strategy of ODDM.
Conventional TFMC modulation pulse/waveform design principles are governed by the WH frame theory, which ensures global (bi)orthogonality across the whole TF domain. For practical systems having a limited bandwidth and frame duration, the MC modulation pulse design only needs to satisfy the local or sufficient (bi)orthogonality inside the TF region of interest.
Then, by reformulating the (bi)orthogonal pulse design problem, we reveal the ``mystery" of DDOP, and justify its hitherto unknown benefits, which are achieved without violating the classic WH frame theory. Finally, we present the salient properties of the ODDM modulation, including its signal localization, bandwidth efficiency, implementation benefits, and its ISAC potentials.

In the conventional TFMC modulation schemes such as OFDM, the orthogonality among OFDM subcarriers facilitates the following benefits, which justify its wide adoption in many communications systems \cite{ofdm,tff,mct,keller_proc_ieee_2000,jiang_proc_ieee_2007,weinstein2009,mc_book_lly_2009}: (A.1) the immunity to frequency-selective multipath effects; (A.2) the low-complexity single-tap equalization; (A.3) the trivial bandwidth partitioning; (A.4) the straightforward adoption to \ac{mimo} systems. Nonetheless, the following OFDM deficiencies are also widely recognized: (D.1) the high \ac{papr} that encumbers the power amplifier design; (D.2) the bandwidth efficiency erosion due to the CP overhead and owing to the unloaded subcarriers inserted at the band edge for controlling the OOBE; (D.3) the sensitivity to CFO that includes Doppler shift and oscillator mismatch; (D.4) the OOBE that affects the coexistence of asynchronous users. Over the past six decades, a variety of transmission and reception techniques have been developed, including \ac{ofdm-im}\cite{ofdm-im} and the recent multi-band \ac{dft-s-ofdm} amalgamating with \ac{im} \cite{hanzo1}, in order to mitigate the OFDM deficiencies. However, they tend to compromise some of the key OFDM benefits.

In the proposed DDMC modulation scheme such as ODDM, modulating information in the DD domain and orthogonality between the ODDM subcarriers with respect to the Doppler resolution and between multiple ODDM symbols with respect to the delay resolution brings about a number of benefits: (A.1) both time- and frequency-diversity can be attained in doubly-selective channels, leading to reliable transmissions in these hostile channels; (A.2) low channel estimation pilot overhead and less frequent channel estimation; (A.3) having a preferable ambiguity function, which is attractive for future ISAC applications; (A.4) reduced CP cost, since only one CP is required for a single data frame, resulting in an improved bandwidth efficiency; (A.5) having a moderate PAPR for an appropriate pulse $u(t)$ and for suitable values of $M$ and $N$.
  {We remark that ODDM does not achieve a common decomposition of arbitrary LTV channels into independent subchannels (as OFDM does for LTI channels). However, because of its orthogonality with respect to the delay and Doppler resolutions of the channel, ODDM better matches the delay and Doppler characteristics of the channel, and it is expected to lead to lower implementation complexity for both communications and sensing applications.}
As a novel and fundamentally new waveform, ODDM or DDMC in general is still in its early stage of development. There are many challenging open questions to be answered in future research. Some of them are listed in the following.

\begin{itemize}
  \item To detect ODDM signals over doubly selective channels, the conventional low-complexity single-tap equalization does not provide satisfactory performance. {Existing OTFS detectors based on message passing \cite{viterbo_twc_2018,ampotfs}, on linear maximal-ratio combining\cite{thaj2020low}, and on minimum mean squared error\cite{mmse_otfs} can be extended to ODDM\cite{oddm_detector}, which can offer good performance, but at a high computational cost.} Receiver designs based on deep-learning to explore the DD domain channel properties and signal structures are also of interest. In this regard, receivers exhibiting performance vs. complexity trade-offs are essential for practical systems.

  \item {Circuit impairments including CFO, DC offset, IQ imbalance, phase noise, etc., constitute a critical issue in practical transceiver designs \cite{rfme}. Due to its orthogonality with respect to the fine JTFR, the ODDM performance erosion under realistic circuit impairments requires further investigation. Also, pilot or signal designs conceived for compensating these impairments in OFDM systems\cite{ofdm_cfo_dc,ofdm_cfo_iq_blind,ofdm_cfo_iq} may be also extended to ODDM systems. }

  \item Similar to conventional OFDM waveform and its relatives, ODDM can also have diverse beneficial variants. For example, ODDM may be further evolved to DFT-S-ODDM, as a bridge between SC transmission and DDMC transmission. Like OFDM-IM\cite{ofdm-im} and MC-CDMA\cite{nathan_ieice_2002,yang_commag_2003}, ODDM can also be combined with error control coding using modern channel codes (like LDPC, Turbo and polar codes), index modulation or conventional CDMA technologies for achieving good bandwidth-/power-/energy-efficiencies.

  \item For ODDM to be applicable to practical systems, its transmitter and receiver must be flexibly scalable both in terms of antennas and users.
          {The implementation architecture of MIMO-ODDM systems, as well as the associated resource allocation strategies for both multi-user systems and multi-cell systems, are relevant future directions.}

  \item Performance analysis in terms of its the achievable rates, error probability and channel code design for doubly-selective channels constitute further research challenges to be tackled.

  \item {To provide flexibility in term of resource allocation and system optimization, it may be of interest to combine ODDM with interference cancellation based strategies, such as \ac{rsma} \cite{rsma} and \ac{noma}\cite{noma} for both \ac{miso} and MIMO systems.}

  \item Security is at essence in wireless systems, where the unique physical layer channel properties of legitimate users can be exploited to provide security to complement upper-layer security relying on secret keys. For ODDM signals, due to the associated spreading of signal, there is an improved grade of physical layer security. This is another compelling research item.

  \item Furthermore, the popular reconfigurable intelligent surfaces can also be combined with ODDM as they are capable of improving the coverage of space–air–ground integrated networks
        in the era of \ac{6g}.

  \item The PAPR of ODDM is expected to be low to moderate when $N$ is small\cite{papr_oddm_vtc25f}. However, a thorough comparison to alternative techniques such as OFDM and OTFS will be interesting. PAPR reduction techniques such as clipping may also be considered.

  \item Note that this paper focuses on the pulse design aspects of DD domain modulation and on the comparisons between the TF and DD modulation in pulse design and its consequences. Whether one ought to adopt a TF modulation scheme or DD modulation scheme is an important question. This will depend on the type of channels, on the required system performance metric, as well as transmitter and receiver complexity, etc.

  \item As an alternative to DDMC, chirp/affine-based designs such as \ac{ocdm} and \ac{afdm} have been recently proposed to address the challenges of LTV channels \cite{chirp_2001,OCDM2016TCOM,YiyinWangWCL2023,haif2024novel,bemani2021afdm,Shen2025ICC,2025_ICC_Chengyang_OCDM}.
        For example, \cite{chirp_2001} proposes an eigenfunction-based transmission scheme over an LTV channel with a linear DD spreading function, which aligns with the channel-dependent waveform design discussed in Section IV. On the other hand, \ac{afdm} employs sequences of chirps, or aliased chirps, to transmit information symbols\cite{bemani2021afdm}. Although chirps were originally introduced in analog form\cite{chirp_2001,OCDM2016TCOM}, recent chirp-based modulation waveforms are represented and evaluated as digital sequences. Their explicit continuous-time waveform representations, which are critical for practical implementation, remain unclear. Future research could investigate whether these chirp/affine-based designs offer additional advantages.

  \item  ISAC is an emerging service in future systems. Although much progress has been reported in the last decade, there are many open research problems to be solved in this research area. On the other hand, while it was only preliminarily investigated in \cite{LuningLinSPL2025}, the application of ODDM to ISAC will also have many fundamental and practical questions to be answered, such as the capacity or achievable rates of the signals over various channels, sensing limits or performance bound, potential trade-offs between communications and sensing, practical considerations including the effect of imperfect frequency {offsets}, timing and frequency synchronizations, etc., on system performance. Solving these problems will help pave the way of practical {applications} of ODDM or DDMC in future ISAC systems.
\end{itemize}

\section*{Acknowledgment}
The authors would also like to thank the Editor and anonymous reviewers for their professionalism, as well as the time and effort they devoted to reviewing our long manuscript. Their constructive comments and suggestions have helped us to improve the writing and readability of the paper. We sincerely thank them.

\appendices
\section{Proof of Proposition 2}
Since the period of $g(t)$ is $\frac{T}{N}$, we have
\begin{align}
  g(t)=g(t+\dot n\frac{T}{N}), 0 \le \dot n \le N-1,
\end{align}
for $0\le  t < \frac{T}{N}$.
Then, bearing in mind that $T=1/\Delta F$, the ambiguity function of $g(t)$ is given by
\begin{align}
  \mathcal A_{g,g}(0,n\Delta F) & =  \int_0^{T_g} g(t)g^*(t)e^{-j2\pi n\Delta Ft} dt, \nonumber                                                                 \\
                                & = \sum_{\dot n=0}^{N-1} \int_{\dot n \frac{T}{N}} ^{(\dot n+1) \frac{T}{N}} g(t)g^*(t)e^{-j2\pi n\Delta Ft} dt, \nonumber     \\
                                & = \sum_{\dot n=0}^{N-1} e^{-j2\pi\frac{n\dot n}{N}} \times \int_0^{\frac{T}{N}} g(t)g^*(t)e^{-j2\pi n\Delta Ft} dt, \nonumber \\
                                & = \delta(n),
\end{align}
for $|n|\le N-1$, {and the last equality is based on the fact that $g(t)$ is normalized to unit energy. This completes the proof}.

\begin{figure}
  \centering
  \includegraphics[width=8.5cm]{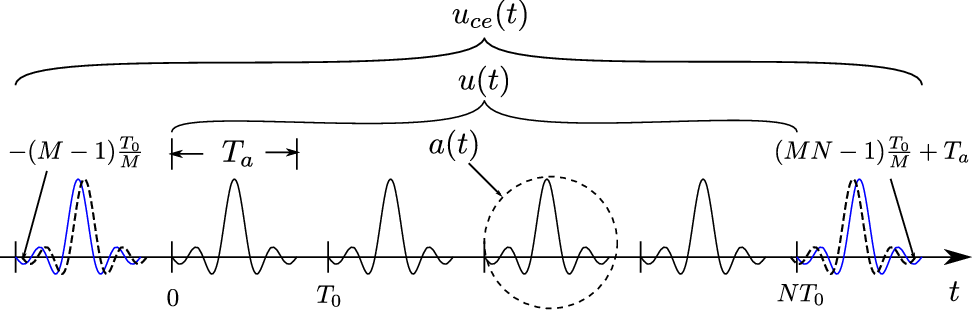}
  \caption{$u_{ce}(t)$ for $D=1$}
  \label{gce1}
\end{figure}

\section{Proof of Proposition 3}
Let us first check the periodicity of $u_{ce}(t)$ within the range of $-(M-1)\frac{T_0}{M}\le t \le (MN-1)\frac{T_0}{M}+T_a$, which corresponds to the start of the first subpulse of $u(t+(M-1)\frac{T_0}{M})$ and the end of the last subpulse of $u(t-(M-1)\frac{T_0}{M})$, respectively. Recall that
\begin{align}\label{gtat}
  u(t)=\sum_{n=0}^{N-1} a(t-nT_0),
\end{align}
we can divide $u(t)$ into $N$ segments, where $u(t) = \sum_{n=0}^{N-1}u_n(t)$ and the $n$th segment is given by
\begin{align}
  u_n(t)  =
  \begin{cases}
    u(t) & nT_0 \le t < (n+1)T_0 \\
    0    & \textrm{otherwise}
  \end{cases}.
\end{align}

Let $D=\lceil T_a/T_0 \rceil$. If $D=1$, we have
\begin{align}
  u_n(t)= a(t-nT_0),
\end{align}
which implies that the periodicity within $-(M-1)\frac{T_0}{M}\le t \le (MN-1)\frac{T_0}{M}+T_a$ can be obtained by cyclically extending $u(t)$ in (\ref{gtat}) to
\begin{align}
  u_{ce}(t)=\sum_{n=-1}^{N} a(t-nT_0).
\end{align}
Similarly, when $D>1$, the periodicity can be obtained by cyclically extending $u(t)$ in (\ref{gtat}) to
\begin{align}\label{ncpcs}
  u_{ce}(t)=\sum_{n=-D}^{N-1+D} a(t-nT_0).
\end{align}
Two examples of $u_{ce}(t)$ associated with $D=1,2$ are shown in Fig. \ref{gce1} and Fig. \ref{gce2}, respectively, where the first subpulse of $u(t+(M-1)\frac{T_0}{M})$ and the last subpulse of $u(t-(M-1)\frac{T_0}{M})$ are also plotted with dashed lines.

\begin{figure}
  \centering
  \includegraphics[width=8cm]{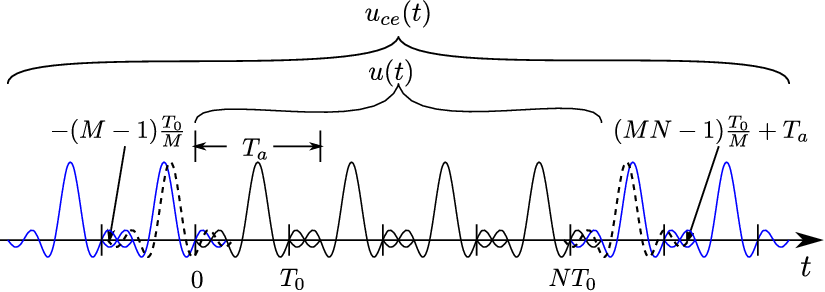}
  \caption{$u_{ce}(t)$ for $D=2$}
  \label{gce2}
\end{figure}

Next, let us verify the ambiguity functions. Due to the aforementioned periodicity of  $u_{ce}(t)$, we have
\begin{align}\label{cycforinte}
  u_{ce}(t)=u_{ce}(t+\dot nT_0), 0\le \dot n \le N-1.
\end{align}
for $m\frac{T_0}{M} \le  t \le m\frac{T_0}{M}+T_u$, where $|m| \le M-1$ and $T_u=(N-1)T_0+T_a$. Then, the cross ambiguity function between $u_{ce}(t)$ and $u(t)$ for $|n|\le N-1$ and $|m|\le M-1$ is given by 
\begin{align}
   & \mathcal A_{u_{ce},u}\left(m\frac{T_0}{M}, n\frac{1}{NT_0}\right) \nonumber                                                                                             \\
   & =  \int_{m\frac{T_0}{M}}^{m\frac{T_0}{M}+T_u} u_{ce}(t)u^*\left(t-m\frac{T_0}{M}\right)e^{-j2\pi n\frac{1}{NT_0}(t-m\frac{T_0}{M})} dt, \nonumber                       \\
   & = \int_0 ^{T_u} u_{ce}\left(t+m\frac{T_0}{M}\right)u^*(t)e^{-j\frac{2\pi n}{NT_0}t} dt, \nonumber                                                                       \\
   & \stackrel{(a)}{=} \sum_{\dot n=0}^{N-1} \int_{\dot n T_0} ^{\dot n T_0+T_a} u_{ce}\left(t+m\frac{T_0}{M}\right)a^*(t-\dot nT_0)e^{-j\frac{2\pi n}{NT_0}t} dt, \nonumber \\
   & \stackrel{(b)}{=} \sum_{\dot n=0}^{N-1} e^{-j2\pi\frac{n\dot n}{N}}  \int_0^{T_a} u_{ce}\left(t+m\frac{T_0}{M}\right)a^*(t)e^{-j\frac{2\pi n}{NT_0}t} dt, \nonumber     \\
   & = \delta(n)\int_0^{T_a} u_{ce}\left(t+m\frac{T_0}{M}\right)a^*(t)dt \nonumber                                                                                           \\
   & =\delta(n)\delta(m), \label{orthoam}
\end{align}
where $\stackrel{(a)}{=}$ and $\stackrel{(b)}{=}$ are due to (\ref{gtat}) and (\ref{cycforinte}), respectively. {This} completes the proof.



%





\ifCLASSOPTIONcaptionsoff
  \newpage
\fi



%
\bibliographystyle{IEEEtran}
\bibliography{oddm}

%




\end{document}